\documentclass[10pt,twocolumn,english,superscriptaddress,lengthcheck,prl]{revtex4-2}
\usepackage[latin9]{inputenc}
\usepackage{color}
\usepackage{babel}
\usepackage{verbatim}
\usepackage{amsmath}
\usepackage{amssymb}
\usepackage{graphicx}
\usepackage{esint}
\usepackage[unicode=true,
bookmarks=false,
breaklinks=false,pdfborder={0 0 1},backref=section,colorlinks=true]
{hyperref}

\makeatletter
\usepackage[caption=false]{subfig}
\hypersetup{colorlinks=true,linkcolor=blue,filecolor=magenta,urlcolor=green}

\renewcommand{\citep}[1]{\cite{#1}}
\renewcommand{\citet}[1]{\cite{#1}}
\newcommand{\jcaption}[1]{\normalsize #1}

\makeatother
\begin{document}
	
	\newcounter{mybibstartvalue}
	\setcounter{mybibstartvalue}{1}
	\usecounter{enumiv}\setcounter{enumiv}{\value{mybibstartvalue}}	
\author{Rauf Giwa, Pavan Hosur}
\affiliation{Department of Physics, University of Houston, Houston 77204, USA}
\title{Fermi arc criterion for surface Majorana modes in superconducting
	time-reversal symmetric Weyl semimetals}
\begin{abstract}
	Many clever routes to Majorana fermions have been discovered by exploiting the interplay between superconductivity and band topology in metals and insulators. However, realizations in semimetals remain less explored. We ask, ``under what conditions do superconductor vortices in time-reversal symmetric Weyl semimetals -- three-dimensional semimetals with only time-reversal symmetry -- trap Majorana fermions on the surface?'' If each constant-$k_{z}$ plane, where $z$ is the vortex axis, contains equal numbers of Weyl nodes of each chirality, we predict a generically gapped vortex and derive a topological invariant $\nu=\pm1$ in terms of the Fermi arc structure that signals the presence or absence of surface Majorana fermions. In contrast, if certain constant-$k_{z}$ planes contain a net chirality of Weyl
	nodes, the vortex is gapless. We analytically calculate $\nu$ within a perturbative scheme and provide numerical support with a lattice model. The criteria survive the presence of other bulk and surface bands  and yield phase transitions between trivial, gapless and topological vortices upon tilting the vortex. We
	propose Li(Fe$_{0.91}$Co$_{0.09}$)As and Fe$_{1+y}$Se$_{0.45}$Te$_{0.55}$ with broken inversion symmetry as candidates for realizing our proposals. 
\end{abstract}
\maketitle


The interplay of band topology and superconductivity
has paved new routes to Majorana fermions (MFs) -- as topologically
protected zero energy bound states trapped in topological defects
such as superconductor vortices  \cite{Alicea2012,Aste2010,Beenakker2013,Elliott2015,Fu2008,Hosur2011,Kitaev2000,Leijnse2012,Liu2018,Liu2017,Lutchyn2011,Ma2017,Mohanta2014,Mourik2012,Nadj-Perge2014,Qi2011,Read2000,Rokhinson2012,Sato2010,Sato2017}.
Following realizations in semiconductor nanowire-superconductor heterostructures
\cite{Stanescu2010,Lutchyn2011,Mourik2012}, MFs were recently found for the first time in a three-dimensional (3D) system -- at the ends
of vortices in the bulk superconductor $\mathrm{FeSe}_{0.45}\mathrm{Te}_{0.55}$ \cite{Xu2016a,Machida:2019wg,Kong:2019we,Wang2018,Zhang2018}. This inspires
a fundamental question: in a 3D superconductor, what
properties of the normal state band structure ensure that
vortices trap protected MFs at their ends? Restricting to bands with time-reversal symmetry ($\mathcal{T}$), since $\mathcal{T}$
enables a Cooper instability to begin with, sufficient conditions
are known in two generic cases: a band insulator (metal) yields MFs if it is topological \cite{Fu2008} (a modestly doped topological insulator \cite{Hosur2011}).

The third type of generic $\mathcal{T}$-preserving band material is a time-reversal symmetric Weyl semimetal (TWSM) \cite{Herring1937,volovik2009universe,Wan2011,Armitage2018}.
Here, point intersections between non-degenerate bands create Weyl
nodes (WNs) that possess a chirality of $\pm1$ and appear in
quadruplets to respect $\mathcal{T}$ and Brillouin zone periodicity.
Weyl semimetals constitute topological matter as they are immune to
perturbations that do not hybridize anti-chiral WNs, i.e., WNs of
opposite chirality and exhibit numerous topological responses \cite{Hosur2013a,Burkov2014,Burkov2018,Burkov_2015,ChenAxionResponse,ZyuninBurkovWeylTheta,Hosur2015c,Hu:2019aa,Juan:2017aa,Kobayashi2018,LandsteinerAnomaly,li20203d,Loganayagam2012,Nagaosa:2020aa,Niemann2017,PhysRevB.92.161110,QiWeylAnomaly,RanQHWeyl,Shekhar:2015aa,Sonowal:2019kyy,SonSpivakWeylAnomaly,Tabert2016,VazifehEMResponse,Wang2017,Wang_2018,WeylDielectric,NielsenABJ}. On the surface, the bulk band topology manifests as Fermi arcs (FAs)
that connect surface projections of anti-chiral WNs.

Motivated by the quest for MFs, we ask, ``what is the fate of a superconductor
vortex in a TWSM?'' We show that there are three possible vortex
phases -- (i) gapped, with end-MFs; (ii) gapped, without end-MFs;
(iii) gapless, with topologically protected chiral MFs dispersing
along the vortex axis $\hat{\mathbf{z}}$. Crucially, we prove that
the vortex phase relies solely on basic band structure data, namely,
the FA configuration on the surface normal to $\hat{\mathbf{z}}$
and the locations of the bulk WNs. Remarkably, simply tilting the
vortex can drive transitions between the three phases. 

\begin{figure}
\includegraphics[width=0.53\columnwidth]{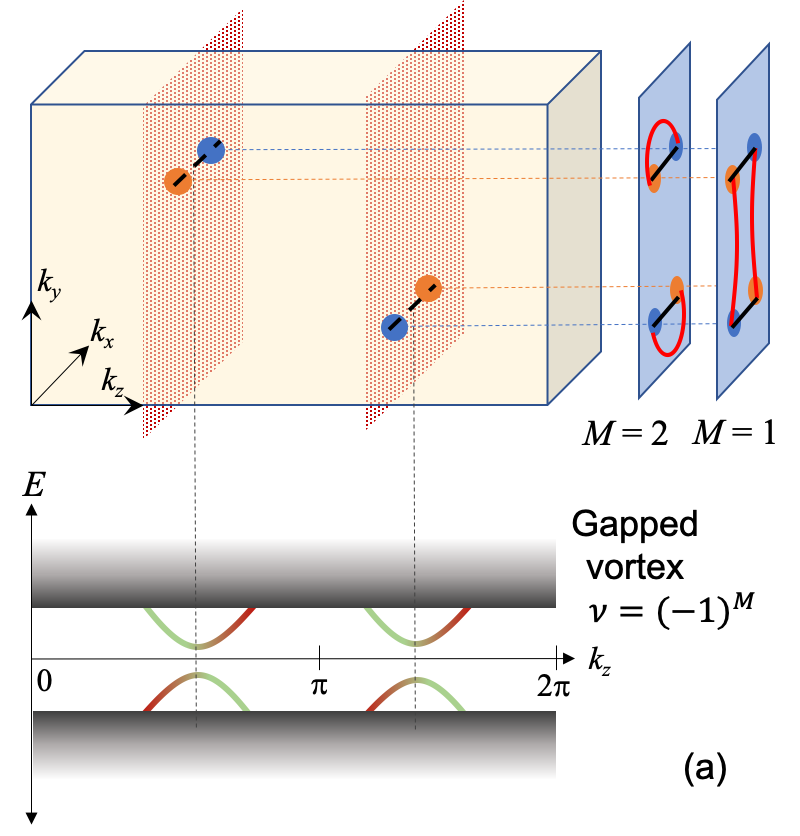}~~\includegraphics[width=0.46\columnwidth]{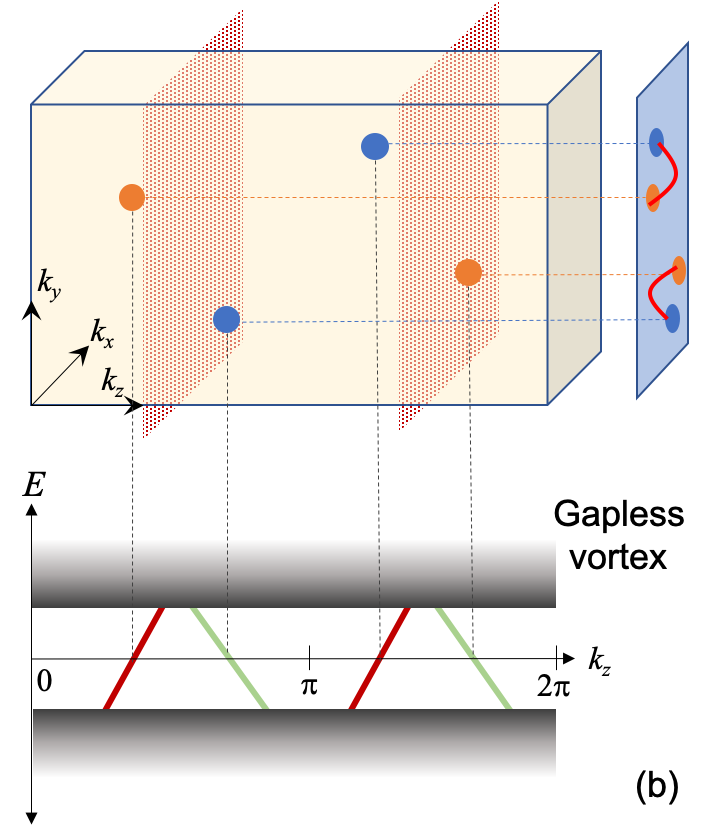}

\caption{\jcaption{Schematic of the main result. Orange (blue) dots denote
right(left)-handed WNs which produce right(left)-moving chiral MFs,
colored red (green), inside the vortex. To determine the vortex phase,
identify pairs of anti-chiral WNs at the same $k_{z}$ and project
the line joining them onto the surface. If these lines (solid black),
along with the FAs (red curves), form $M$ closed loops, the vortex
is gapped and has a topological invariant $\nu=(-1)^{M}$ (a), whereas
open arcs produce a gapless vortex (b).}\label{fig:Schematic}}
\end{figure}

The criteria for the phases are as follows (see Fig.  \ref{fig:Schematic}).
Within each constant-$k_{z}$ plane, identify the pair(s) of anti-chiral
WNs that are closest to each other in periodic $\boldsymbol{k}$-space.
Connect the partners with a geodesic and project it onto the surface.
From the remaining WNs, identify the next closest pair(s) and project
their geodesic(s) onto the surface, and so on for all WNs and constant-$k_{z}$
planes. If all the WNs find partners in this process, the surface
Brillouin zone will contain a set of lines that, along with the FAs,
form closed \emph{Fermi-geodesic loops}. In general, the
surface will also carry closed Fermi loops and Dirac nodes. If the
total number of Fermi-geodesic loops, Fermi loops and Dirac nodes is $M$, we predict a gapped vortex with a topological invariant
\begin{equation}
\nu=(-1)^{M}\label{eq:topological invariant}
\end{equation}
Thus, odd $M$ yields a topologically protected MF in the vortex core
on the surface whereas even $M$ does not. All WNs find partners only
if each constant-$k_{z}$ plane contains equal numbers of left- and
right-handed WNs. For a minimal TWSM with four WNs at $(\pm\boldsymbol{K}_{1},\pm\boldsymbol{K}_{2})$,
a vortex direction such that $\left|K_{1z}\right|=\left|K_{2z}\right|$
ensures this, while general TWSMs with more WNs need a mirror or glide
symmetry plane parallel to $\hat{\mathbf{z}}$ -- which is present
in most TWSMs \cite{Meng2019,Weng2015,Niemann2017,Liu2014b} --
to ensure all WNs are partnered. Generically, though, some WNs will lack partners
and the surface will host open \emph{Fermi-geodesic arcs}
whose end-points will be projections of the \foreignlanguage{american}{unpartnered}
WNs. Each such WN will contribute a 1D chiral MF to the bulk vortex
spectrum with a chirality equal to its own chirality times the vorticity
of $\pm1$. The vortex will be in a gapless phase protected by $k_{z}$-conservation,
analogous to the topological protection of a Weyl semimetal by 3D
momentum conservation. 

These criteria hold for general pairing symmetries provided the
superconductor is gapped in the absence of a vortex. They survive
doping around the WNs if the resulting Fermi surfaces are well-separated and the presence of trivial Fermi surfaces with rare exceptions. They are also immune to surface effects unless the surface is exposed to a topological insulator, in which case $M$ effectively acquires the odd number of surface Fermi loops or Dirac nodes of the latter. Finally, (\ref{eq:topological invariant}) captures the known results for metals \cite{Hosur2011}
and insulators \cite{Fu2008}, which lack Fermi-geodesic contours but may have Fermi loops and Dirac nodes.

Eq. (\ref{eq:topological invariant}) is obtained
 by computing the $\mathbb{Z}_{2}$ topological invariant for the vortex viewed as a 1D superconductor \cite{Ryu2010,Schnyder2009,Kitaev2000}.
We require a mild assumption in the clean limit: for a given
WN, if the two nearest anti-chiral WNs at the same $k_{z}$
are at distances $\Delta K_{1}$ and $\Delta K_{2}$, 
then $e^{\frac{\hbar \xi}{\Delta_0}[v_1(\Delta K_{1})^{2}-v_2(\Delta K_{2})^{2}]}\gg 1$
or $\Delta K_{1}\gtrsim\Delta K_{2}$, where $\Delta_0$ and $\xi$ are superconducting properties -- the uniform pairing amplitude and coherence length, respectively -- and $v_i$ is the typical Weyl velocity along $\Delta K_i$. This condition ensures that the dominant
hybridization is between chiral MFs from neighboring anti-chiral
WNs, assuming hybridization is driven by band curvature. If hybridization
is due to non-magnetic disorder that is smooth over distance $\ell_{D}$,
the requirement becomes $\Delta K_{1}\gg\ell_{D}^{-1}\gg\Delta K_{2}$
while magnetic disorder invalidates (\ref{eq:topological invariant}). Disorder can be suppressed in principle
whereas band curvature is unavoidable, so a physical regime of validity of (\ref{eq:topological invariant}) exists. We neglect hybridization between equi-chiral chiral MFs, i.e., chiral MFs of the same chirality, which amounts to smoothly deforming any accidentally gapless non-chiral vortex mini-bands away from zero energy.

Heuristically,
(\ref{eq:topological invariant}) says that vortex-end MFs are present
(absent) if the TWSM normal state is ``closer'' to a topological
(trivial) insulator. To see this, imagine
moving the WNs along the geodesics and
annihilating them in pairs. If all WNs get annihilated,
the resulting insulator will be topological (trivial) if the surface
FAs evolve into an odd (even) number of surface Fermi loops, while
the vortex will be topological (trivial). However,
the vortex spectrum remains gapped in the process, so its topological
state before and after WN-annihilation must be the same. 

Recently, Refs. \cite{Konig2019,Qin2019} showed that $s$-wave superconductor
vortices in Dirac semimetals can trap helical MFs protected by crystal
symmetries. Unlike those vortices, the gapless vortex here is a 1D phase of matter rather than
a critical point as it cannot be gapped out by perturbatively changing
the crystal space group. Ref. \cite{Yan2019} found that below
a critical doping in a lattice model of a Dirac semimetal with two
Dirac nodes, an $s$-wave vortex normal to the line
joining the nodes is gapped and traps a surface MF. The MF survives
when the Dirac semimetal is perturbed into a minimal TWSM with four
WNs in the plane normal to the vortex, assuming $s$-wave pairing
even with broken inversion symmetry ($\mathcal{I}$) in the TWSM.
In comparison, our criteria include the TWSM results of Ref. \cite{Yan2019}
at low doping, but allow arbitrary numbers and configurations
of type-I WNs, trivial Fermi surfaces and filled topological bands
in the bulk, FAs, Fermi loops and Dirac nodes on the surface,
and arbitrary pairing that opens a full gap when uniform.

\emph{Continuum analytics:-} Consider a canonical WN
of chirality $h=\pm1$ described by $H_{W}(\boldsymbol{P})=h\sum_{j}v_{j}\Sigma_{j}P_{j}-\mu$,
where $\Sigma_{j}$, $j=X,Y,Z$, are Pauli matrices spanning the lowest two bands
and $\boldsymbol{P}$ is the momentum relative to the WN. At $P_{Z}=0$,
$H_{W}$ resembles the 2D surface Hamiltonian of a 3D topological
insulator \cite{Beenakker2013,Hosur2011,Fu2008,Burkov2011c}. If $v_X=v_Y$ and $\mu=0$, it yields a pseudospin-polarized MF with $\langle\Sigma_Z\rangle=w$ in the core of an $s$-wave superconductor vortex, $\Delta(\boldsymbol{R})=\Delta(R)e^{iw\Theta}$, $w=\pm1$ \cite{Fu2008}. 
Being topologically protected, the MF will survive albeit with
partial polarization, $0<w\langle \Sigma_Z\rangle <1$, when $v_X\neq v_Y$, $\mu\neq0$ and the pairing is arbitrary
but real and non-zero on the Fermi surface. In fact, the MF only requires a Fermi surface Berry phase
of $\pi$ in the weak-pairing, smooth-vortex limit \cite{Hosur2011}. When $P_{Z}\neq0$, the MF disperses as $E_{h}=hv_{Z}P_{Z}\langle \Sigma_{z}\rangle $,
thus realizing a chiral MF with chirality $h$ at $P_{Z}=0$ or the $k_{z}$ of the parent WN.
In a real TWSM, $k_{z}$-conservation forbids hybridization between
chiral MFs whose
parent WNs are at different $k_{z}$, resulting in a gapless vortex
{[}Fig. \ref{fig:Schematic}(b){]}.

Next, consider a minimal TWSM with one quadruplet of WNs at $(\pm\boldsymbol{K}_{1},0)$,
$(\pm\boldsymbol{K}_{2},0)$, where WNs at $\pm\boldsymbol{K}_{n}$
are related by $\mathcal{T}$ and have chirality $(-1)^{n}$. Suppose
FAs on the $z=0$ surface connect $\boldsymbol{K}_{1}$ to $\boldsymbol{K}_{2}$
and $-\boldsymbol{K}_{1}$ to $-\boldsymbol{K}_{2}$, and Fermi surfaces
around the WNs are well-separated. In the presence of a superconductor
vortex along $\hat{\mathbf{z}}$, each WN produces a chiral MF dispersing
along $(-1)^{n}\hat{\mathbf{z}}$ with a wavefunction $\psi_{\pm n}=e^{i\boldsymbol{K}_{n}\cdot\boldsymbol{r}}\varphi_{n}(\boldsymbol{r})$,
where $\varphi_{n}(\boldsymbol{r})$ is the zero mode of the vortex
Hamiltonian near the $n^{th}$ WN. The chiral MFs remain robust when
$|\boldsymbol{K}_{n}\xi|\to\infty$, but hybridize for finite $|\boldsymbol{K}_{n}\xi|$.
Neglecting hybridization between equi-chiral chiral MFs, a generic perturbation $H'$ in the basis $\left(\psi_{+1},\psi_{-1},\psi_{+2},\psi_{-2}\right)^{T}$
has the form $H'=\left(\begin{array}{cc}
0 & iQ\\
-iQ^{\dagger} & 0
\end{array}\right)$ where $Q=\left(\begin{array}{cc}
q_{12} & q_{1\bar{2}}\\
q_{\bar{1}2} & q_{\bar{1}\bar{2}}
\end{array}\right)$ and $q_{mn}=\left\langle \psi_{m}\left|H'\right|\psi_{n}\right\rangle $.
If $H'$ preserves $\mathcal{T}$, then $q_{mn}=q_{\bar{m}\bar{n}}$ and
the vortex is a gapped 1D superconductor with topological invariant $\nu=\text{sgn}(\text{Pf}[H'])=\text{sgn}(|q_{12}|^{2}-|q_{1\bar{2}}|^{2})$ \cite{Kitaev2000}. For a spatially
smooth perturbation, $q_{mn}$ decays with $|\boldsymbol{K}_{m}-\boldsymbol{K}_{n}|$;
for instance, band curvature terms yield $q_{mn}\sim e^{-\frac{1}{2}|K_{m}-K_{n}|^{2}\xi/\Delta_{0}}$
for a linear vortex profile with slope $\Delta_{0}/\xi$ \cite{giwa2020fermisuppl}.
Then, $|\boldsymbol{K}_{1}-\boldsymbol{K}_{2}|\lesssim|\boldsymbol{K}_{1}+\boldsymbol{K}_{2}|$
produces a trivial vortex while $|\boldsymbol{K}_{1}-\boldsymbol{K}_{2}|\gtrsim|\boldsymbol{K}_{1}+\boldsymbol{K}_{2}|$
yields a topological vortex with end MFs. On the surface, geodesics
connecting $\boldsymbol{K}_{1}$ to $\boldsymbol{K}_{2}$ and $-\boldsymbol{K}_{1}$
to $-\boldsymbol{K}_{2}$, along with the FAs, form $M=2$ loops.
In contrast, geodesics connecting $\boldsymbol{K}_{1}$ to $-\boldsymbol{K}_{2}$
and $-\boldsymbol{K}_{1}$ to $\boldsymbol{K}_{2}$ form $M=1$ loops
with the FAs. Thus, there is a one-to-one correspondence between $\nu$
and $M$ that is captured by (\ref{eq:topological invariant}). The
Gaussian form of $q_{mn}$ further ensures only logarithmic corrections
to the above inequalities due to $\mathcal{O}(1)$ pre-factors.

Next, consider moving the WNs away from $k_{z}=0$ in pairs while
preserving $\mathcal{T}$ in the normal state. If $K_{1z}=K_{2z}$,
the chiral MF $\psi_{+1}(\boldsymbol{r})$ can hybridize with
$\psi_{+2}(\boldsymbol{r})$ but not with $\psi_{-2}(\boldsymbol{r})$,
so the resulting vortex is adiabatically connected to the trivial vortex where
all WNs are at $k_{z}=0$, $q_{12}\neq0$ and $q_{1\bar{2}}=0$. In contrast,
if $K_{1z}=-K_{2z}$, the adiabatic equivalent with all WNs at $k_{z}=0$
has $q_{1\bar{2}}\neq0$ and $q_{12}=0$, which is a topological vortex. These conclusions extend straightforwardly to more quadruplets of WNs, thus proving (\ref{eq:topological invariant})
for arbitrary configurations of WNs and FAs.

Finally, let us consider the effects of additional bands on our criteria. A filled topological band will produce a 2D Dirac node or Fermi loop on the surface in the normal state, thereby changing the parity of $M$. However, the superconductor vortex will trap another surface MF and acquire a bulk gap due to this band, thus preserving (\ref{eq:topological invariant}). Suppose
the bulk also contains trivial Fermi surfaces, i.e., Fermi surface
that do not enclose WNs or other band intersections. For pairing that is real and non-zero on the Fermi surface, a vortex will contain bands $\varepsilon_{n}(k_{z})\sim\frac{\Delta_{0}}{\xi l_{F}(k_{z})}\left(n+\frac{1}{2}+\frac{\phi_{F}(k_{z})}{2\pi}\right)$
for each trivial Fermi surface, where $n\in\mathbb{Z}$ and $l_{F}(k_{z})$
($\phi_{F}(k_{z})$) is the perimeter (Berry phase) of the Fermi surface
cross-section at $k_{z}$ \cite{Hosur2011}. Unless $\phi_{F}(k_{z})=0$,
trivial Fermi surfaces come in pairs with opposite Berry phases
$\pm\phi_{F}(k_{z})$ to ensure that the constant-$k_{z}$ slice is
a sensible 2D metal. As a result, slices with $\phi_{F}(k_{z})\neq\pi$ will
acquire a vortex minigap $\sim\Delta_{0}/\xi l_{F}(k_{z})$ while slices with $\phi_{F}(k_{z})=\pm\pi$, where
the minigap vanishes, will host counter-propagating chiral MFs that will generically hybridize and gap out. In summary, additional bands will not interconvert gapped and gapless vortices and filled bands always preserve our general criteria.

For gapped vortices, predicted to obey (\ref{eq:topological invariant}), curable violations due to trivial Fermi surfaces occur in two cases: (i) a surface Fermi loop or Dirac node
gets buried under the surface projection of a bulk Fermi surface,
resulting in $M_{\textrm{observable}}=M-1$. This can be cured in
principle by doping to expose the relevant surface state; (ii) a
Fermi surface centered at $k_{z}=0$ or $\pi$ is past the doping
threshold for a vortex phase transition \cite{Hosur2011}, which
changes the vortex phase without changing $M$.
This can be fixed by noting that such a Fermi surface has a pair of
slices away from $k_{z}=0,\pi$ where $\phi_{F}(k_{z})=\pi$. An incurable
violation occurs when $\phi_{F}(k_{z})=\pi$ at the precise $k_{z}$
where a WN exists, the WN is closer to the trivial Fermi surface
than to other anti-chiral WNs at the same $k_{z}$ \emph{and} the
chiral MFs produced by the trivial Fermi surface and the WN have opposite chiralities. Then, $H^{\prime}$ must include hybridization between these MFs, but (\ref{eq:topological invariant}) is ruined because $M$ stays unchanged.

\begin{figure}[h]
\includegraphics[width=0.75\linewidth]{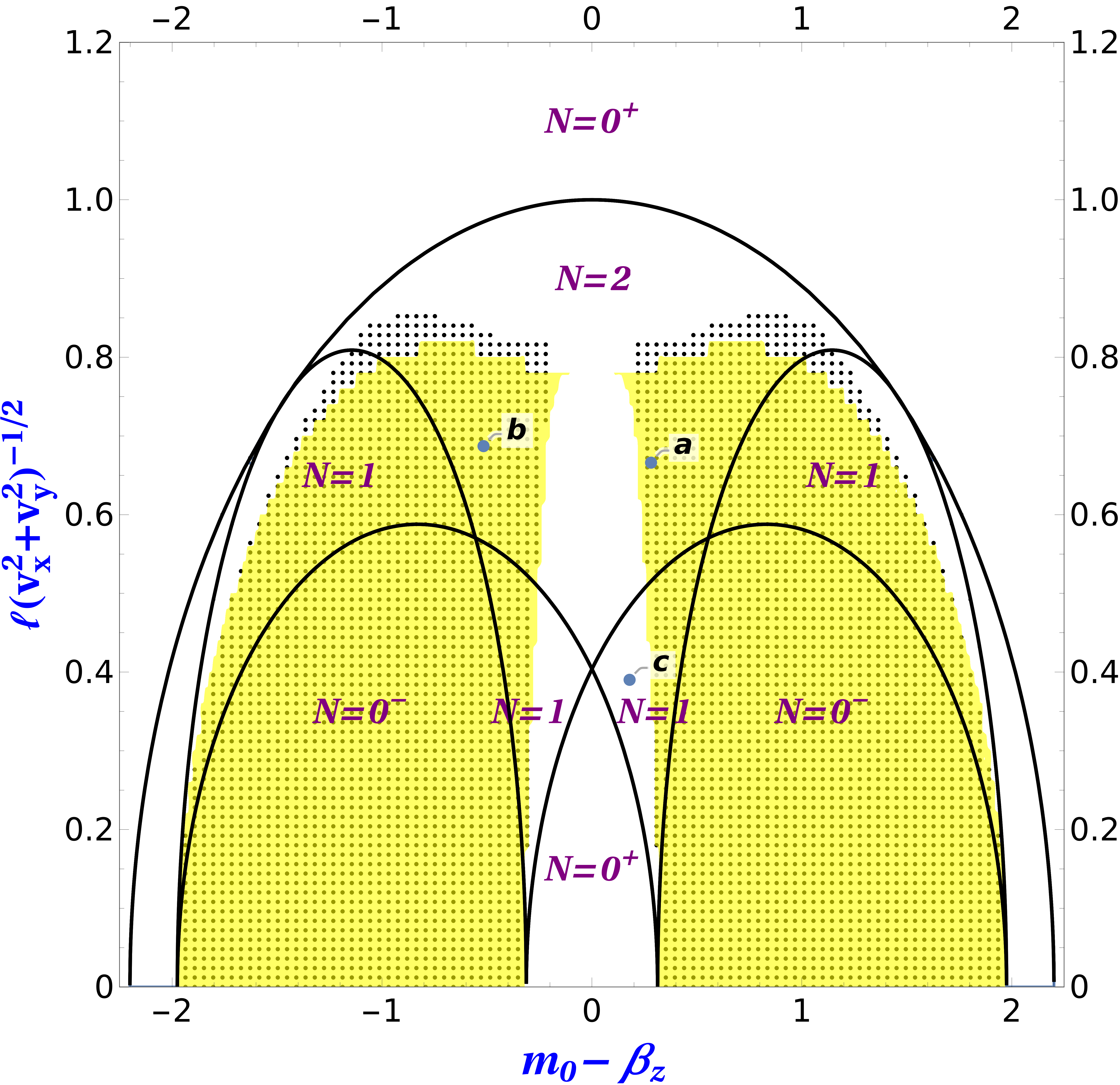}

\caption{\label{fig:Phase diagram}\jcaption{Vortex topological
phases predicted by (\ref{eq:topological invariant}) and computed numerically. The yellow mask (black dots) denote a vortex predicted (computed) to be topological. Black lines separate
the normal states: TWSMs with $N=1,2$ quadruplets of WNs at $k_z=0$ and
trivial/topological ($N=0^{\pm}$) insulators. We fix
$v_{x,y}=1.18,.856$, $\beta_{x,y,z}=.856,1.178,3.0$, $\Delta(r)=0.42\tanh\left(0.3r\right)$
and $L_{x,y}=31$. Points \textbf{t}, \textbf{m} and \textbf{b} are
further studied in Fig. \ref{fig:Straight} }}
\end{figure}

\emph{Lattice numerics:-} We support our general claims with numerics
on an orthorhombic lattice model defined by $H(\boldsymbol{k})=\tau_{x}\boldsymbol{\sigma}\cdot\boldsymbol{d}\left(\boldsymbol{k}\right)+\tau_{z}m\left(\boldsymbol{k}\right)-\tau_{y}\sigma_{z}\ell$
where $d_{i}=v_{i}\sin k_{i}$, $i=x,y,z$, $m(\boldsymbol{k})=m_{0}-\sum_{i}\beta_{i}\cos k_{i}$
and $\tau_{i}$ ($\sigma_{i}$) are Pauli matrices in orbital (spin)
space. Varying $\beta_{x,y,z}$ and $\ell$ allows tuning into trivial
and topological insulating phases as well as TWSMs with up to two quadruplets of WNs each
 at $k_{z}=0,\pi$ at the Fermi level. We then introduce an $s$-wave superconductor vortex
$\Delta(\boldsymbol{r})=\left|\Delta(r)\right|e^{i\theta}$ and
diagonalize the Bogoliubov-deGennes Hamiltonian in 2D real
space at fixed $k_{z}$ to obtain the spectrum and the topological invariant
\cite{Kitaev2000} for a class D 1D superconductor \cite{Schnyder2009,Ryu2010}.
In \cite{giwa2020fermisuppl}, we describe graphical methods for determining the normal state
of $H(\boldsymbol{k})$ and FA-configuration as well as further details
of the lattice numerics. Fig. \ref{fig:Phase diagram} shows that the prediction using
(\ref{eq:topological invariant}) agrees excellently with the
explicit calculation. We found smaller mismatch for larger 
systems or weaker pairing, suggesting that it is
due to departure from the thermodynamic and weak-pairing
limits.

\begin{figure}
\includegraphics[width=0.49\linewidth]{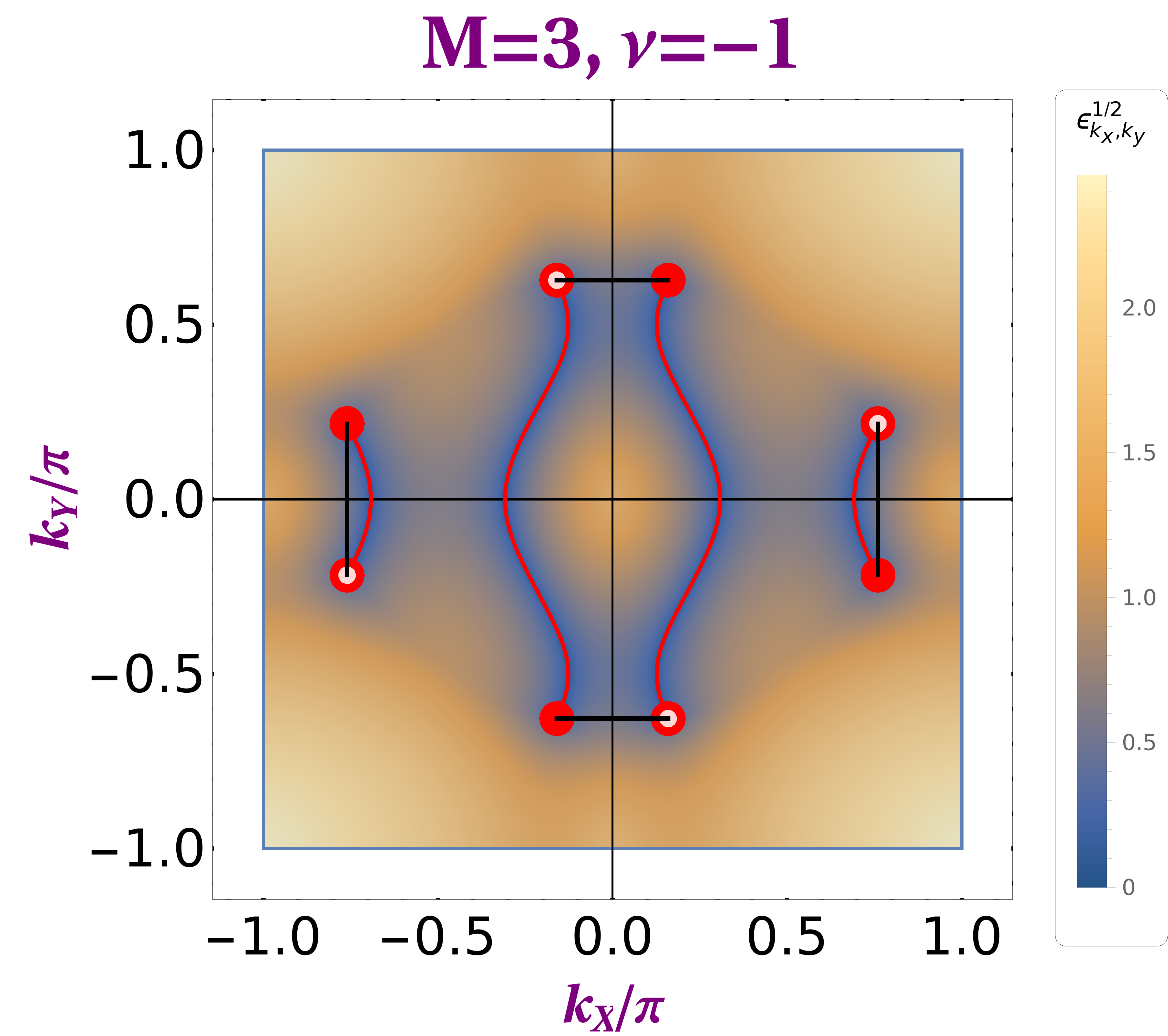}~~\includegraphics[width=0.175\paperwidth,height=0.175\paperwidth]{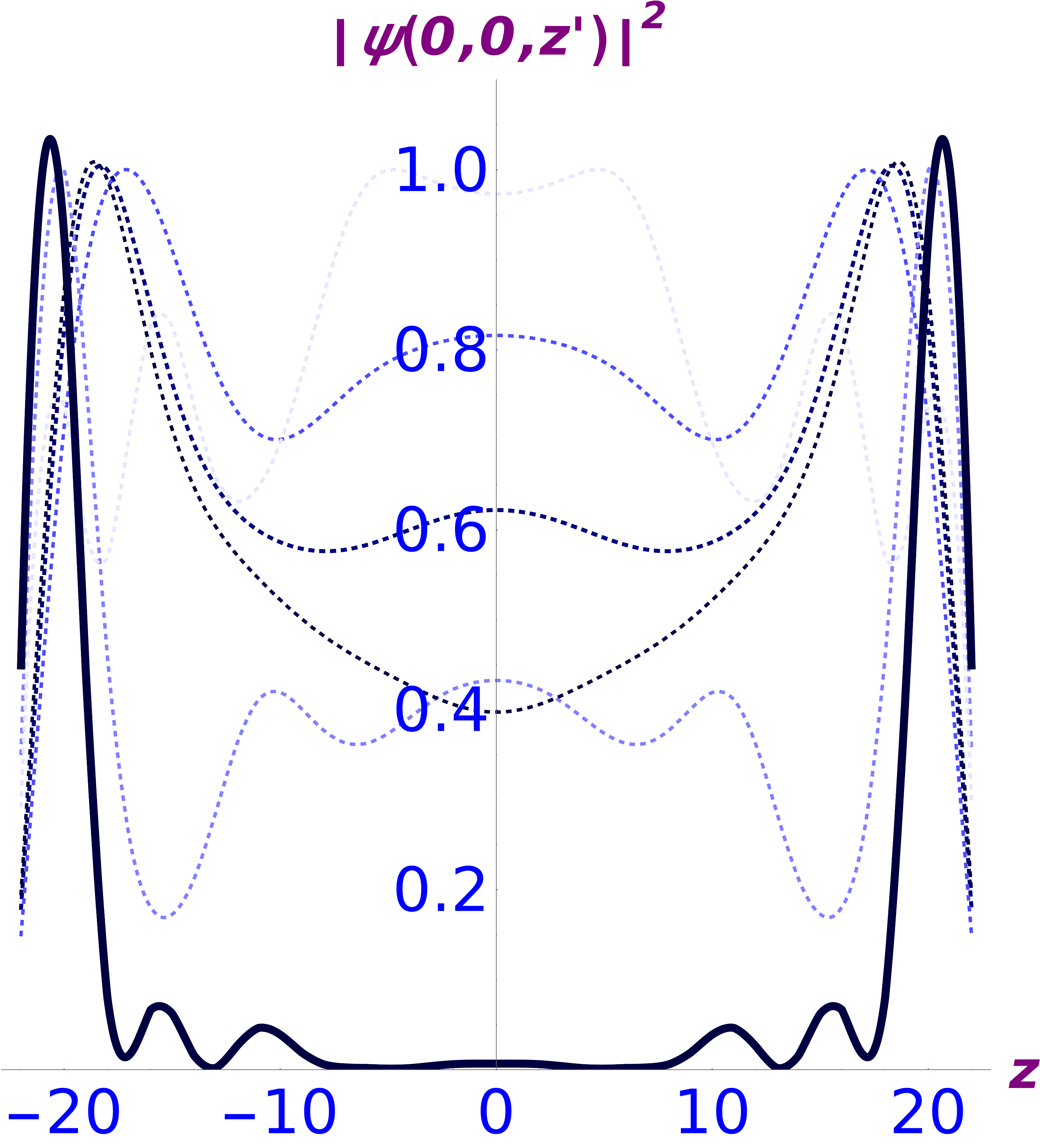}

\rule[0.5ex]{1\columnwidth}{0.5pt}

\includegraphics[width=0.49\linewidth]{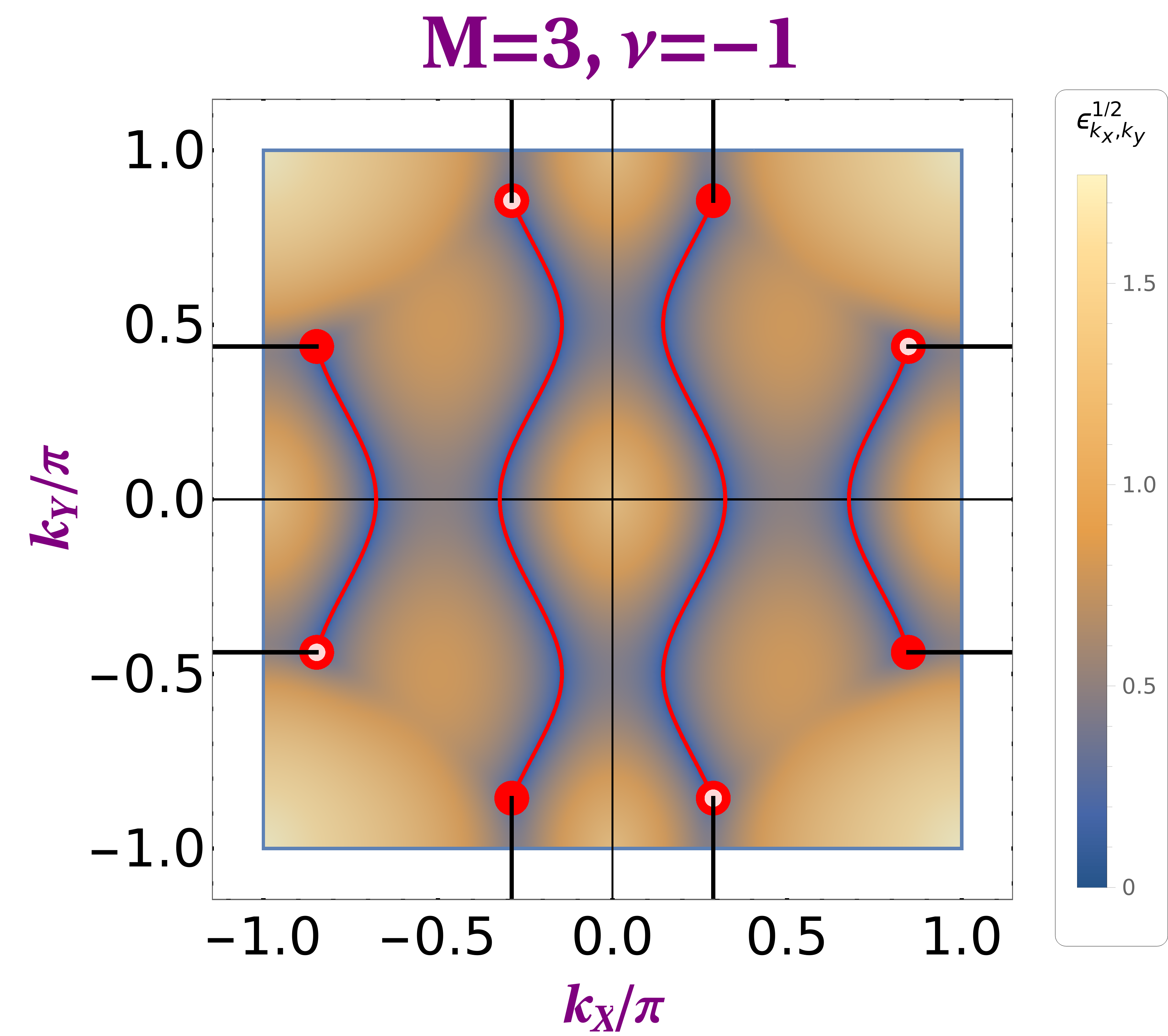}~~\includegraphics[width=0.175\paperwidth,height=0.175\paperwidth]{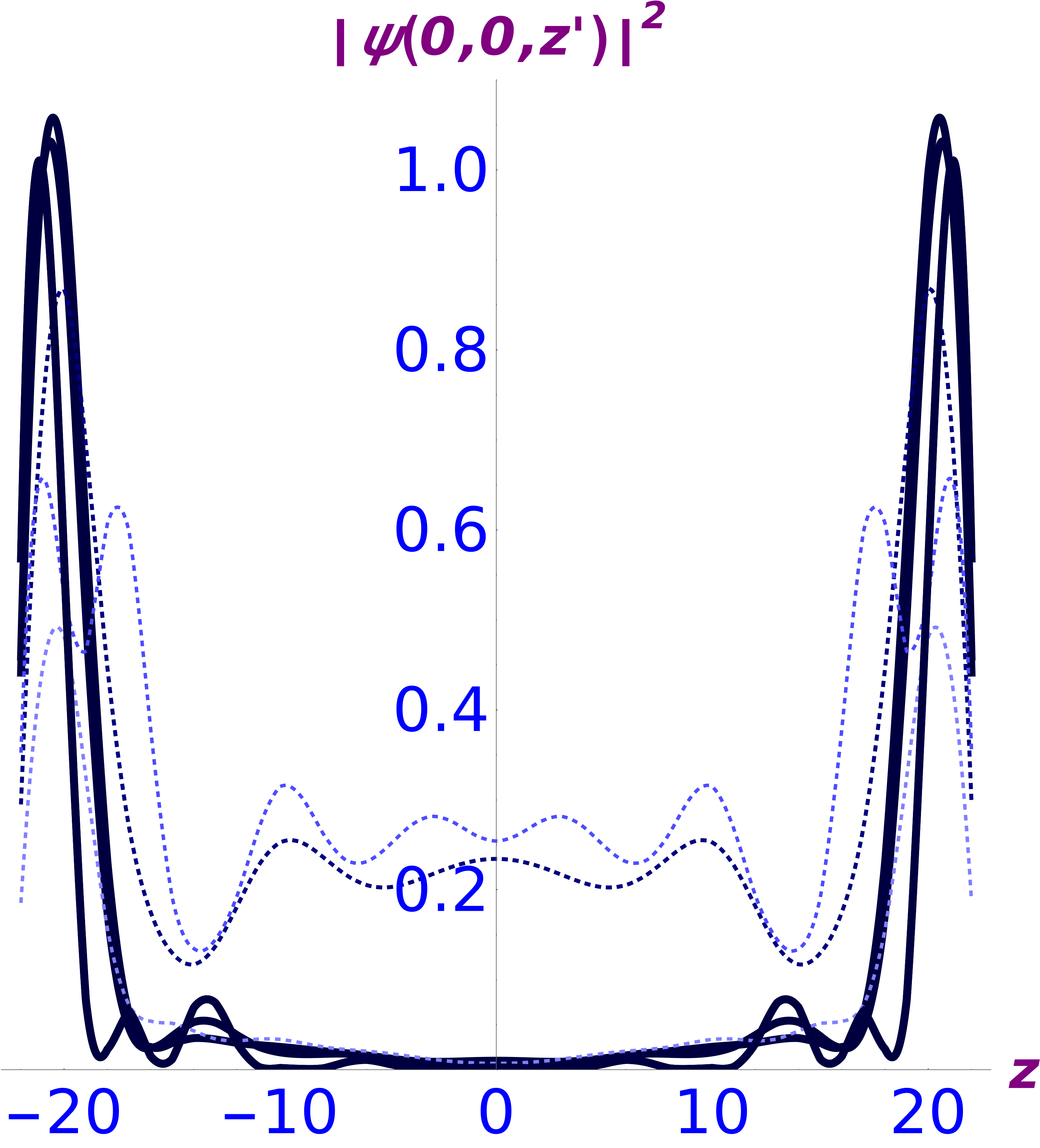}

\rule[0.5ex]{1\columnwidth}{0.5pt}

\includegraphics[width=0.49\linewidth]{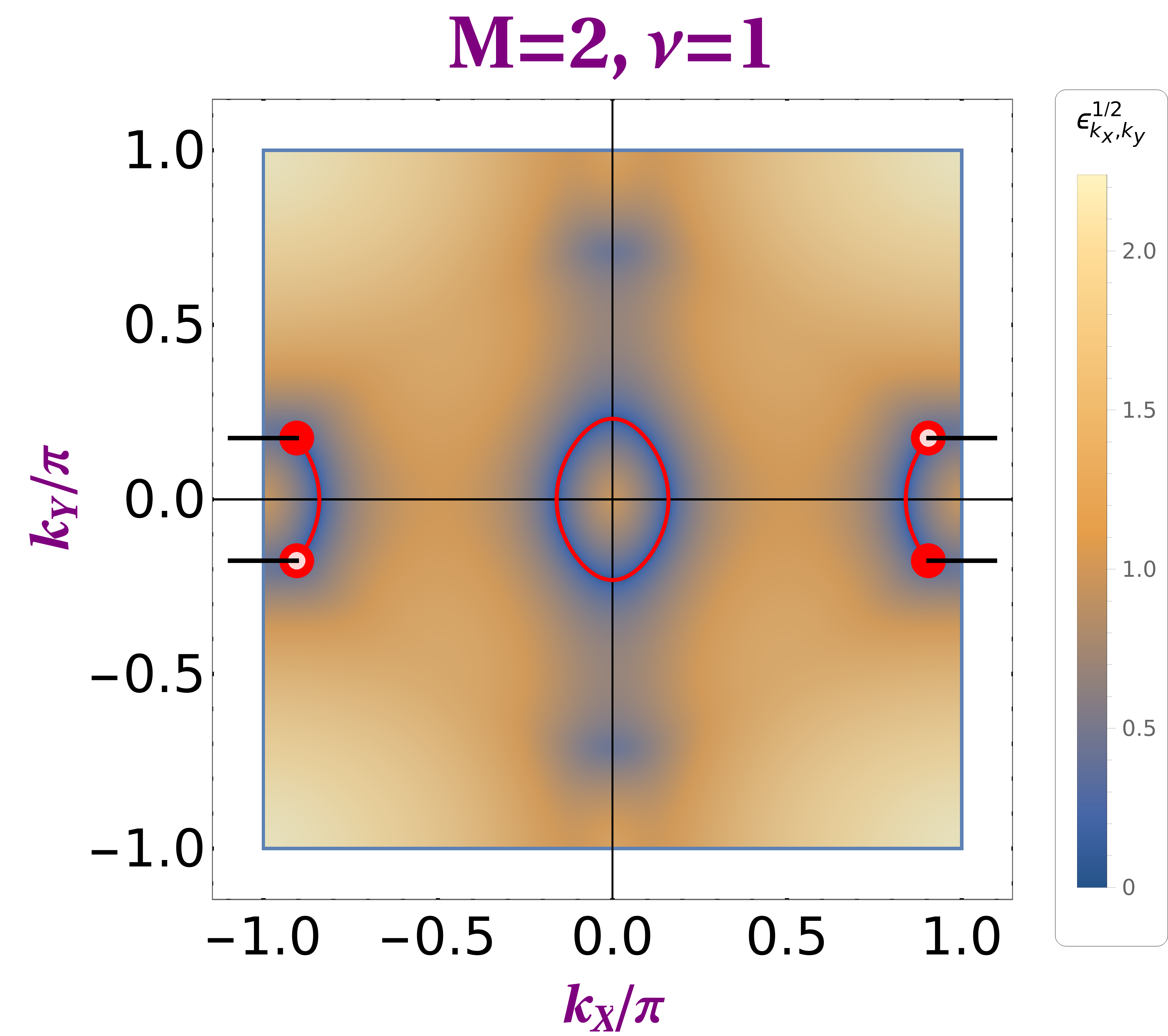}~~\includegraphics[width=0.175\paperwidth,height=0.175\paperwidth]{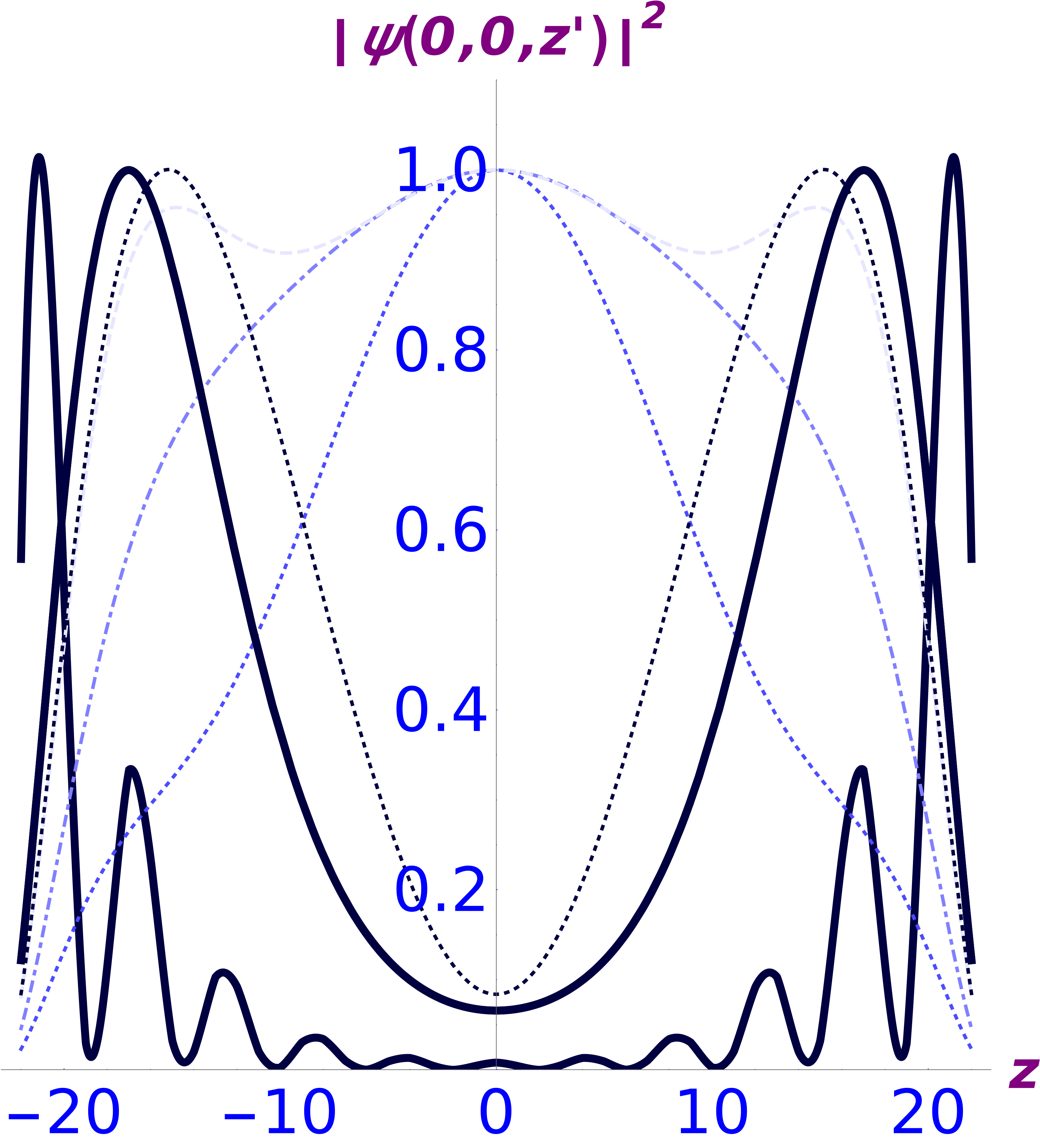}

\caption{\jcaption{Left column: Color plots of the lowest band for a 45-layer slab in the normal state for the points \textbf{t},\textbf{
m} and \textbf{b} in Fig. \ref{fig:Phase diagram} defined by $l=0.942,m_{0}=6.28$
(top), $l=0.972$, $m_{0}=5.48$ (middle) and $l=0.552$, $m_{0}=6.18$
(bottom). Red filled (empty) circles denote surface projections of
right-(left-)handed WNs. Surface projections of geodesics (black lines)
connecting anti-chiral WNs at $k_{z}=0$ form $M$ closed loops along
with the FAs (red curves). Right column: Probability densities of
six lowest energy states along a $z$-oriented vortex
in a $31\times$$31\times$$45$-site system. Bold and dotted lines
mark states with energies $E<5.0\times10^{-3}$, considered "zero energy", and $E>1.0\times10^{-2}$.
} \label{fig:Straight}}
\end{figure}

\begin{figure}
\includegraphics[width=0.45\columnwidth]{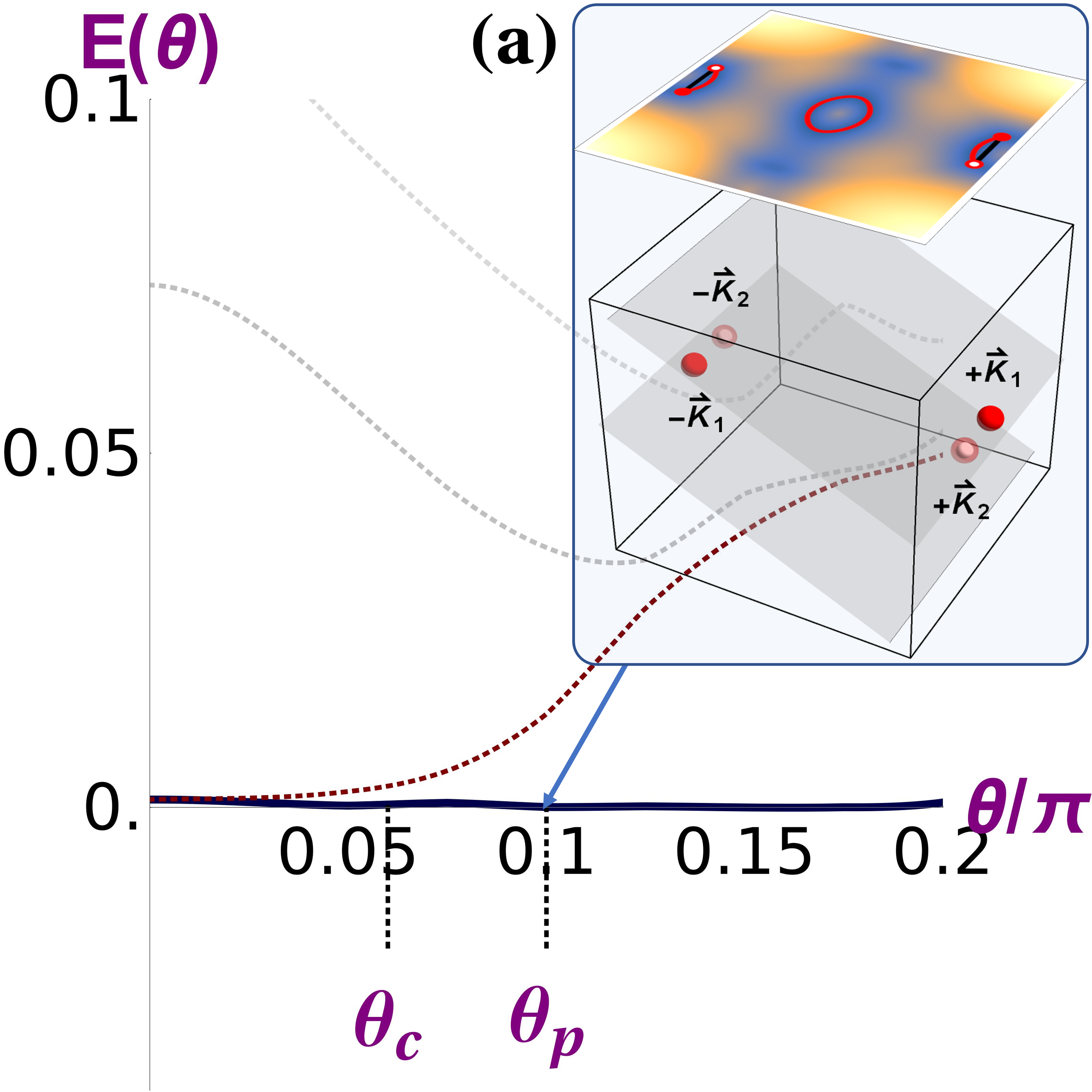} \includegraphics[width=0.45\columnwidth]{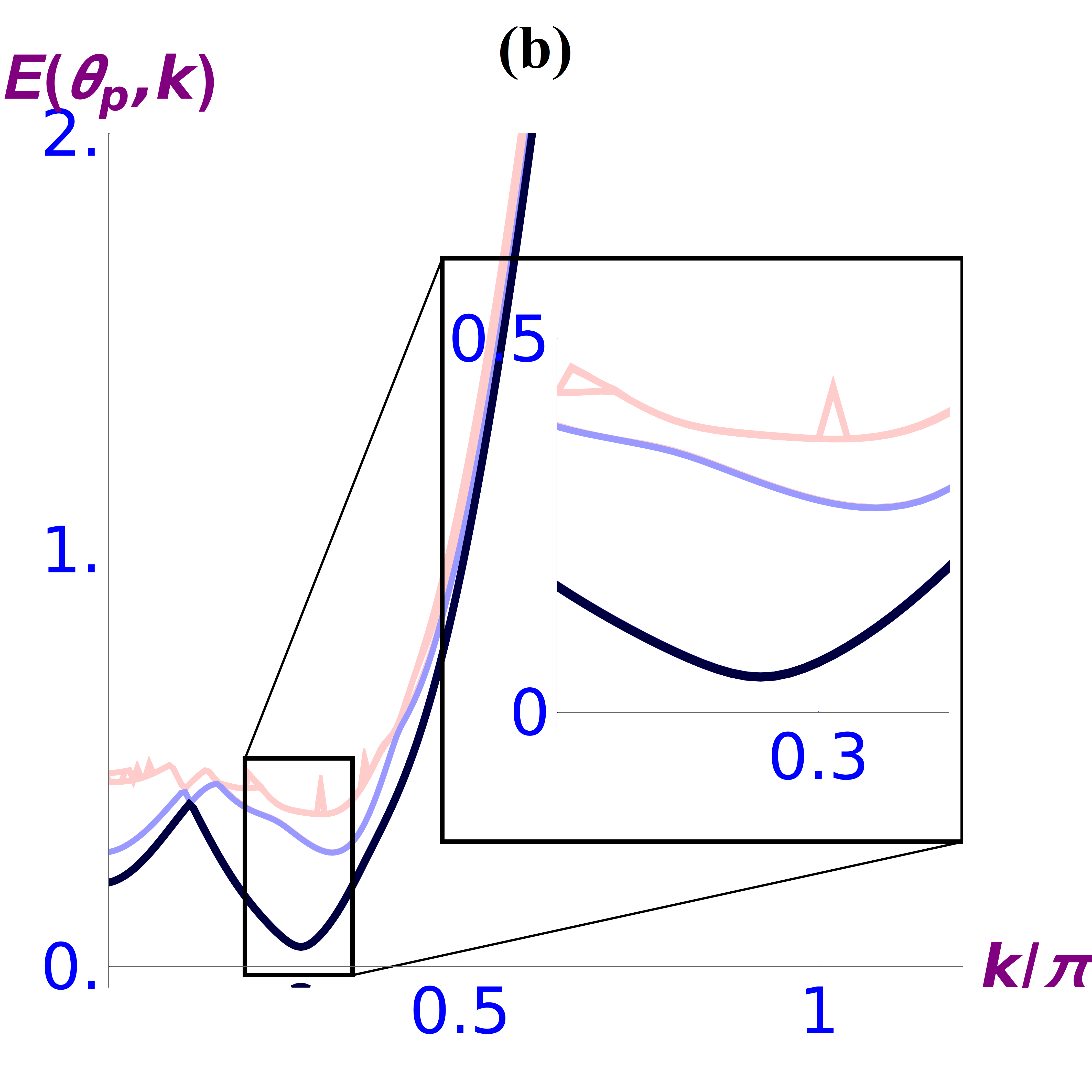}

\includegraphics[width=0.45\columnwidth]{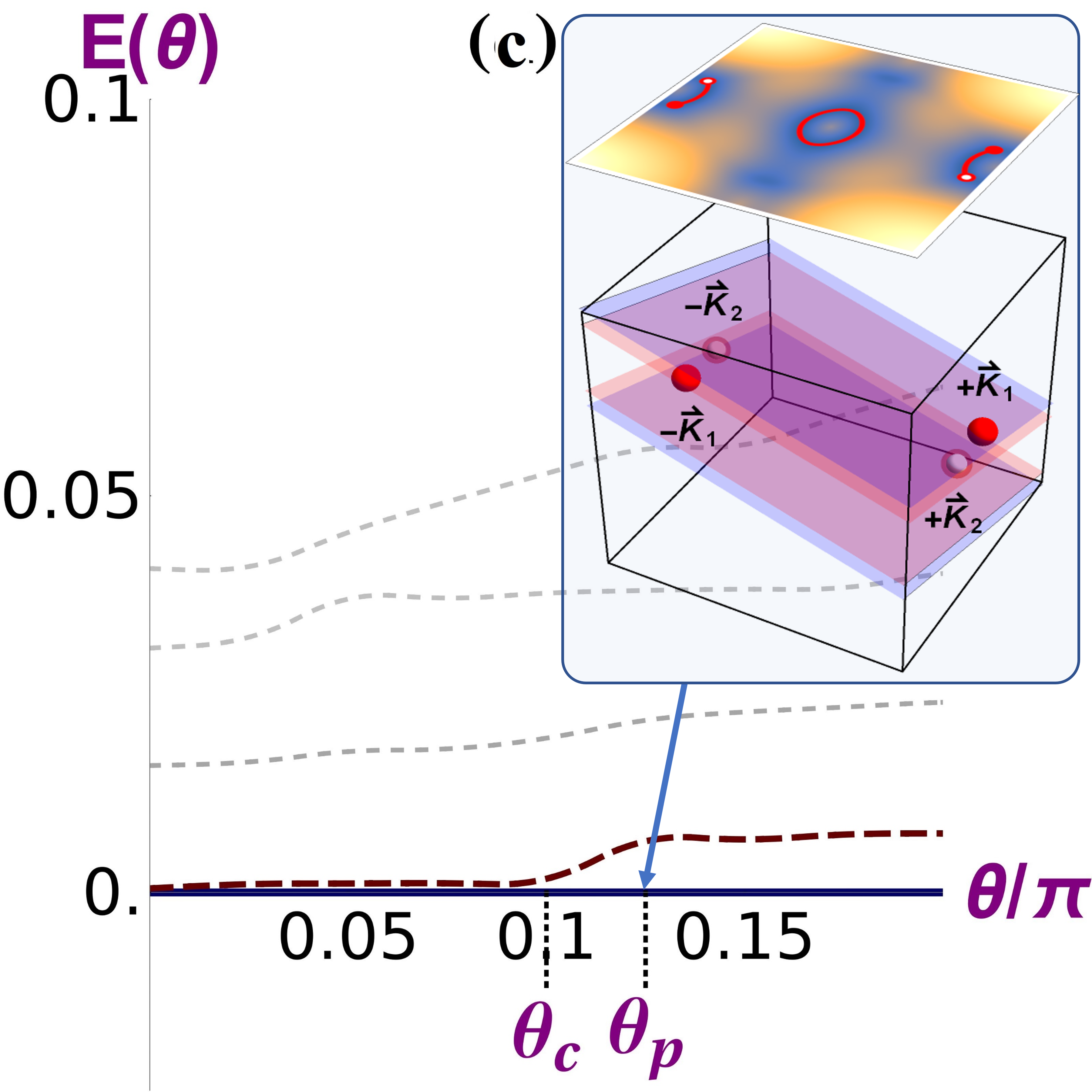} \includegraphics[width=0.45\columnwidth]{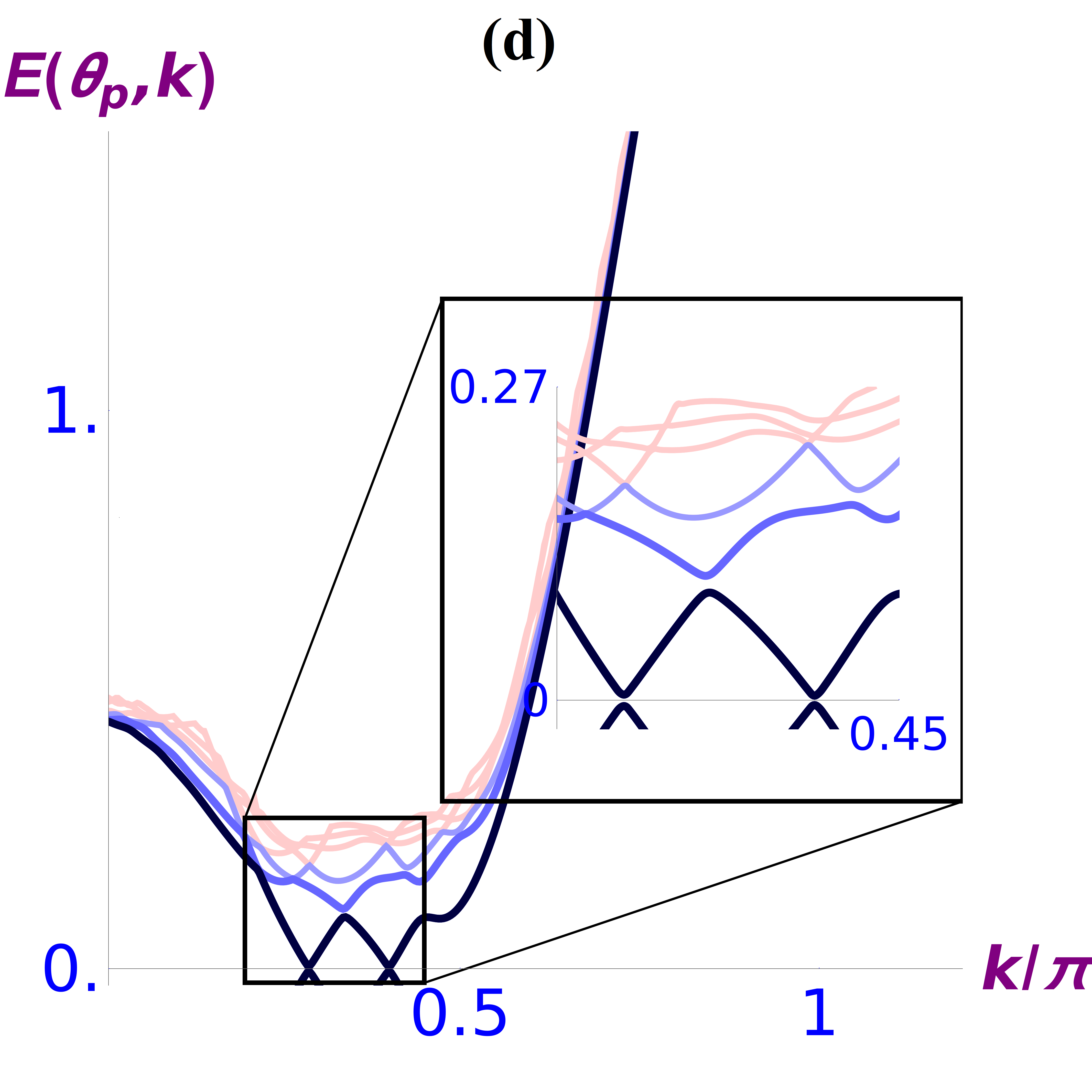}

\caption{\jcaption{Topological phase transitions upon tilting the $M=2$ trivial
vortex in Fig. \ref{fig:Straight} (bottom) obtained by diagonalization
in real-space (a, c) and $k$-space (b, d). (a) Tilting about the
$x$-axis produces $M=3$ (inset) and gaps out one of the two MFs at $\theta_{c}\approx0.06\pi$. (b) At $\theta_{p}=0.1\pi$, the bulk vortex is gapped, so the surviving MF in (a) at $\theta>\theta_{c}$ is protected and the vortex is topological. (c, d) Analogous figures for tilting about $x+y=z=0$. In (c), a small gap opens for one MF at $\theta_{c}\approx0.1\pi$ while the surface has open Fermi-geodesic arcs (inset). (d) The bulk vortex is gapless, indicating that the gap in (c) is a finite size
gap.}\label{fig:Tilted}}
\end{figure}

Fig.  \ref{fig:Straight} shows
the Fermi-geodesic loops in the normal state and the probability
density of the lowest few vortex modes for select points in Fig.
\ref{fig:Phase diagram}. The FAs are obtained by plotting the lowest
energy at each surface momentum in the normal phase and the geodesics
are simply straight lines connecting proximate anti-chiral WNs at
$k_{z}=0$. The probability densities are computed by diagonalizing
the vortex Hamiltonian in 3D real space. For each selected point, we find that 
the number of MFs localized to the vortex ends equals $M$. In Fig. \ref{fig:Tilted}, we show that tilting the vortex drives phase transitions between trivial, topological and gapless vortices since geodesics connect WNs at the same $k_z$. The transitions are expected at infinitesimal tilting in the weak-pairing, smooth-vortex limit, while the numerics find the transitions to occur at small angles.

\emph{Candidate material:-} We propose Li(Fe$_{0.91}$Co$_{0.09}$)As
and Fe$_{1+y}$Se$_{0.45}$Te$_{0.55}$ with broken $\mathcal{I}$
as candidate materials. Li(Fe$_{0.91}$Co$_{0.09}$)As
is a Dirac semimetal with two Dirac nodes on the $c$-axis of the
crystal \cite{Zhang:2019aa} and shows strongly type-II superconductivity below $T_{c}\approx9K$ \cite{Dai2015}. FeSe$_{0.45}$Te$_{0.55}$ realizes
a doped topological insulator that turns into a type-II superconductor
below $T_{c}\approx14.5K$ \cite{Wang2018,Zhang2018}, but the normal
state also has two Dirac nodes along the $c$-axis $\sim15meV$
above the Fermi level that may be accessed with naturally occurring
Fe-dopants \cite{Zhang:2019aa}. Perturbatively breaking
$\mathcal{I}$ while preserving $\mathcal{T}$ will turn the Dirac
semimetals into a TWSM with four WNs at $\pm\boldsymbol{K}_{1},\pm\boldsymbol{K}_{2}$
with $K_{1}^{c}\approx K_{2}^{c}\gg|\boldsymbol{K}_{1}-\boldsymbol{K}_{2}|$.
If superconductivity survives $\mathcal{I}$-breaking, a vortex along
$\hat{\mathbf{z}}$ will be topological (trivial) according to (\ref{eq:topological invariant})
if $\left|K_{1z}\right|=\left|K_{2z}\right|$ and $(\hat{\mathbf{z}}\times\boldsymbol{K}_{1})\cdot(\hat{\mathbf{z}}\times\boldsymbol{K}_{2})>0$ ($<0$), whereas a
vortex in any other direction will be gapless. Assuming typical values
$v\sim10^{5}m/s$ for the Dirac velocity, chemical potential $\mu\sim100K$
relative to the WNs, $\Delta_{0}\sim5K,$ $\xi\sim5nm\ll$ the penetration
depth $d\sim10^{2}nm$ observed in LiFeAs \cite{Qin2019} which guarantees
negligible intervortex tunneling ($\propto e^{-d/\xi}$), and $|\boldsymbol{K}_{1}-\boldsymbol{K}_{2}|/K_{1}^{c}\approx0.1$, we estimate a vortex gap of
$\sim0.1K$. However, the gap depends exponentially on $\xi$, $|\boldsymbol{K}_{1}-\boldsymbol{K}_{2}|$
and $\Delta_{0}$, so it will change substantially for small changes
in their values \cite{giwa2020fermisuppl}.
\begin{acknowledgments}
We acknowledge financial support from the Department of Physics and
the College of Natural Sciences and Mathematics, University of Houston
and NSF-DMR-2047193. 
\end{acknowledgments}


%
%

	\clearpage
	\newpage
	
	\setcounter{mybibstartvalue}{1}	
	\usecounter{enumiv}\setcounter{enumiv}{\value{mybibstartvalue}}	

\appendix\onecolumngrid

\section{\label{A1}Orthorhombic lattice model of a T-WSM\label{sec:Orthorhombic-lattice-model}}

\begin{figure*}
\begin{centering}
\subfloat[]{\includegraphics[width=0.4\linewidth]{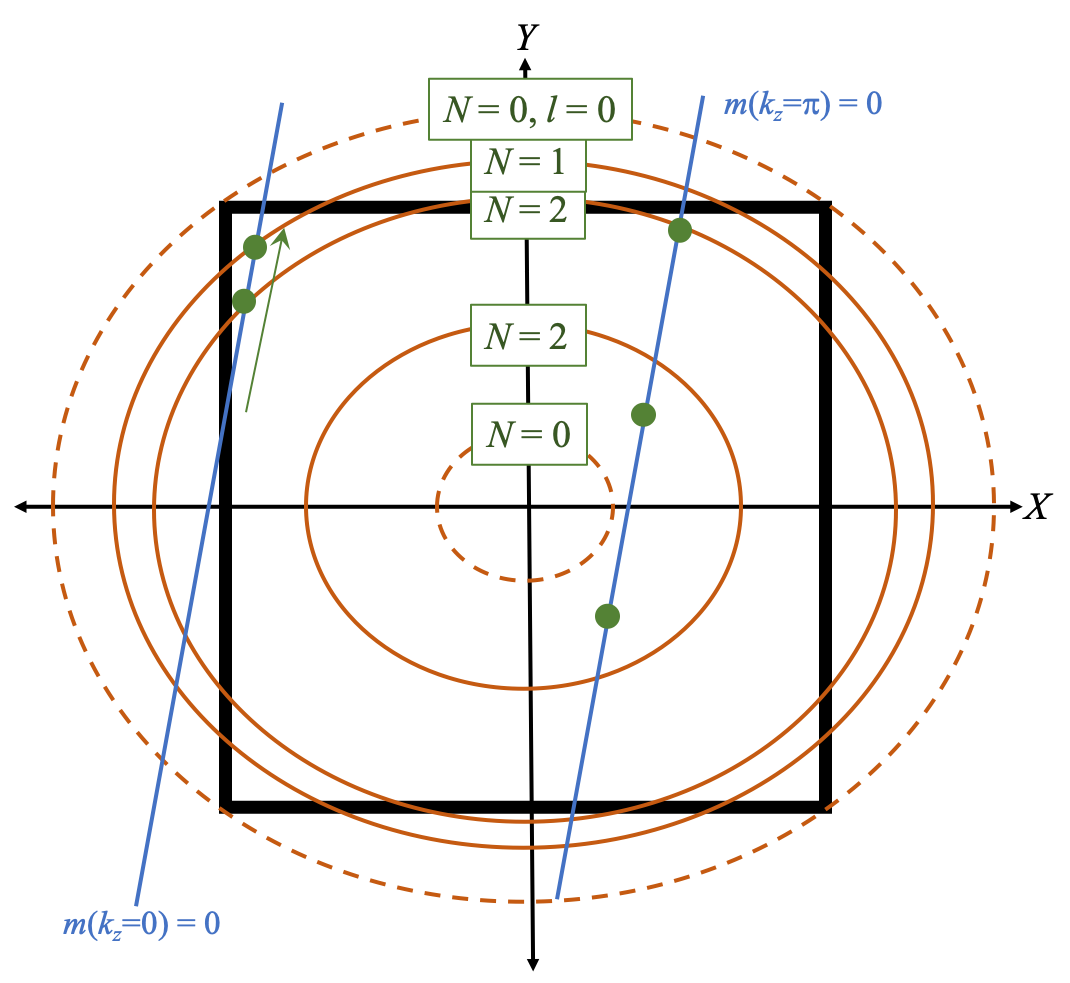}

}\subfloat[]{\includegraphics[width=0.4\linewidth]{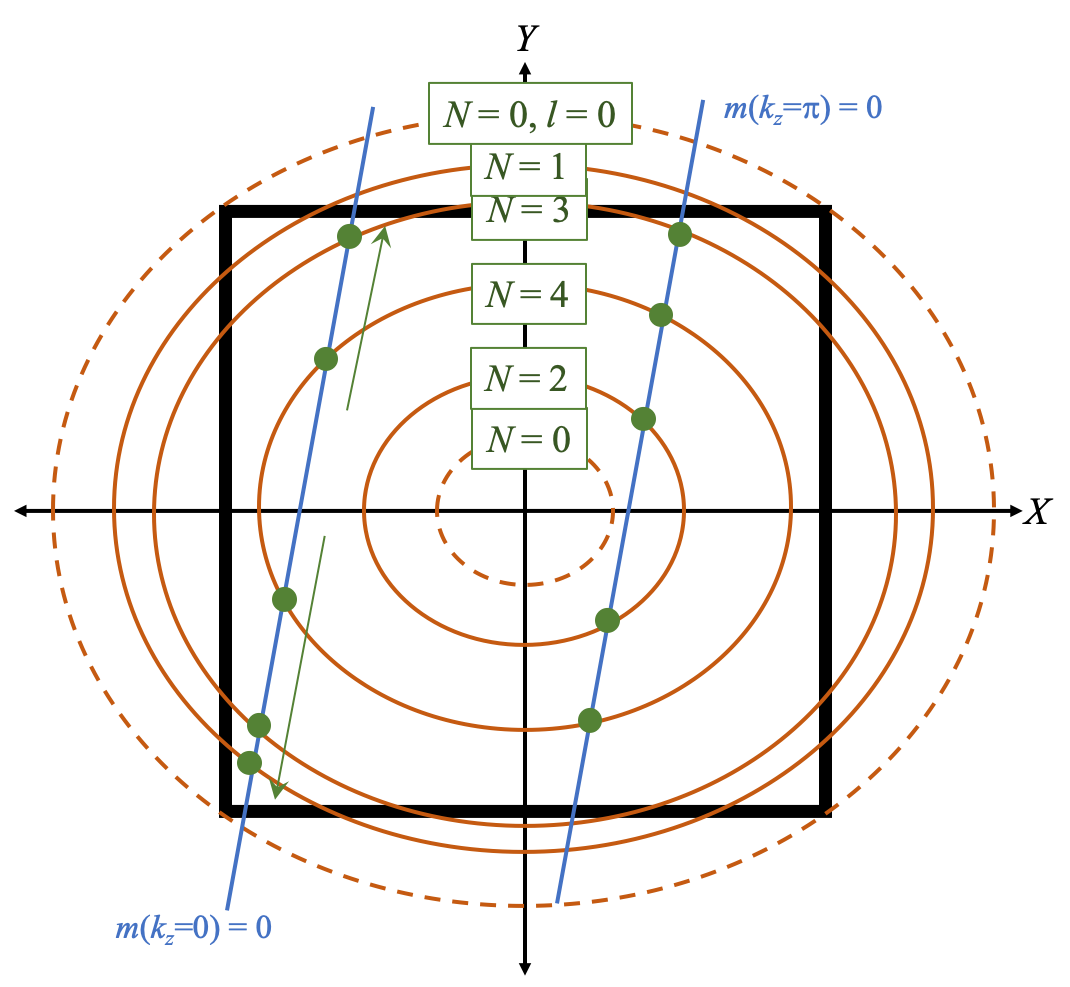}

}
\par\end{centering}
\begin{centering}
\subfloat[]{\includegraphics[width=0.4\linewidth]{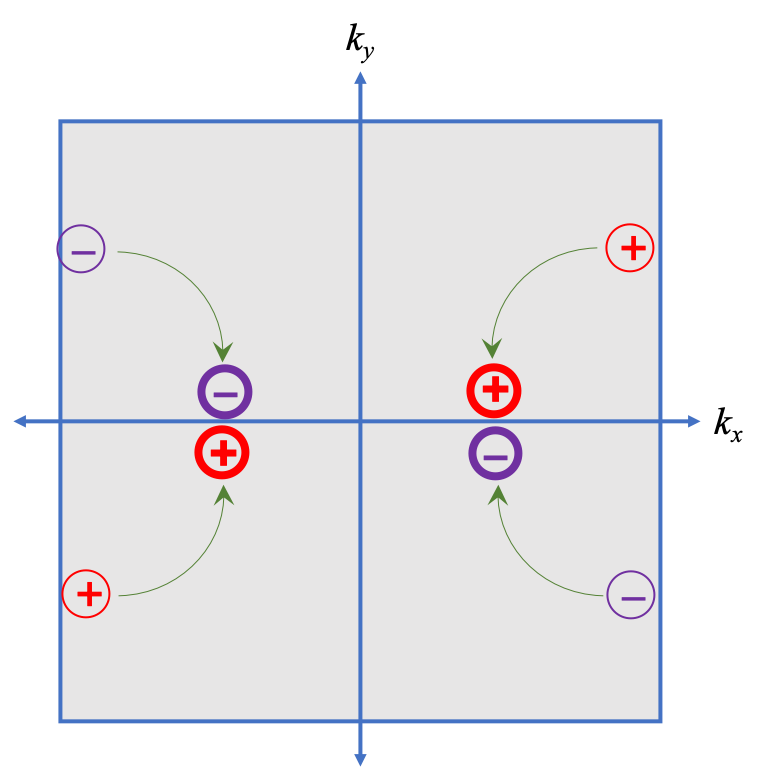}

}\subfloat[]{\includegraphics[width=0.4\linewidth]{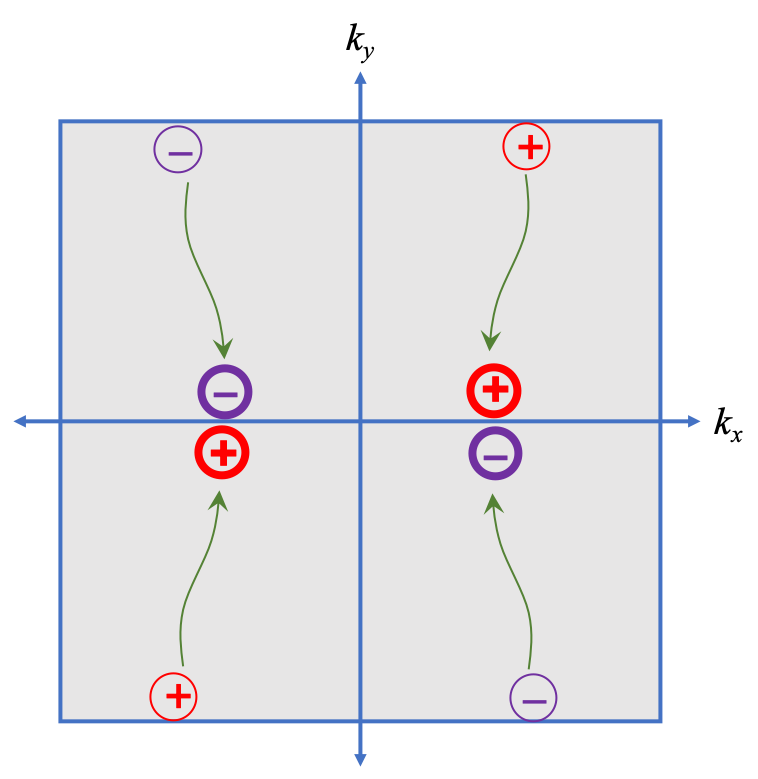}

}
\par\end{centering}
\caption{\jcaption{Prescription to determine the number of WN quadruplets
($N$), the Fermi arc structure on the surface and the $\mathbb{Z}_{2}$
invariant in the insulating phase in the lattice model (\ref{eq:lattice model-1}).
Top: $X=\cos k_{z}$, $Y=\cos k_{y}$ and the ellipses and lines are
given by (\ref{eq:Ellipse-1}) and (\ref{eq:Lines-1}), respectively,
with smaller $|\ell|$ defining larger ellipses. Each ellipse-line
intersection within the defines a quadruplet of WNs in the plane defining
the line. Green arrows indicate the path of the intersections as the
ellipse is enlarged. Solid (dashed) ellipses denote T-WSMs with $N$
quadruplets (insulators with $N=0$). The $\ell=0$ ellipse circumscribes
the square and defines an $\mathcal{I}$-preserving insulator with
$\mathbb{Z}_{2}$ indices given by the parity criterion \cite{Fu2007}.
It has the opposite (same) strong index as the innermost ellipse if
exactly one line (no or both lines) intersects a vertical and a horizontal
edge of the unit square, as shown on the left (right). Bottom: Brillouin
zone of the (001) surface and the effect of moving WNs along the paths
indicated in the top panel on the Fermi arcs. For simplicity, only
the effect of WNs in the $k_{z}=0$ plane is shown; the effects of
$k_{z}=\pi$ WNs are identical. Circles with $\pm$ denote the surface
projections of right/left-handed WNs, and their trajectories as the
ellipse in the top panel is enlarged are indicated by green arrows.
These trajectories trace out the Fermi arcs. If a quadruplet is created
at a $k_{x}=-k_{x}$ plane and annihilated on a $k_{y}=-k_{y}$ plane
or vice-versa, the Fermi arcs close into a single Fermi surface, implying
a change of the bulk strong $\mathbb{Z}_{2}$ topological index. If
a quadruplet is created and destroyed on a $k_{x}=-k_{x}$ (or $k_{y}=-k_{y}$)
plane, the $\mathbb{Z}_{2}$ invariants corresponding to the ellipse
shrunk to a point and the ellipse circumscribing the unit square are
the same.}\label{fig:Ellipse-Line-FermiArc-1}}
\end{figure*}

In this section, we analyze the orthorhombic lattice model studied
in the main text and describe how to determine its topological nature
in the normal state. To recapitulate, the Bloch Hamiltonian is 
\begin{equation}
H(\boldsymbol{k})=\tau_{x}\boldsymbol{\sigma}\cdot\boldsymbol{d}\left(\boldsymbol{k}\right)+\tau_{z}m\left(\boldsymbol{k}\right)-\tau_{y}\sigma_{z}\ell-\mu\label{eq:lattice model-1}
\end{equation}
where $d_{i}=v_{i}\sin k_{i}$, $i=x,y,z$, $m(\boldsymbol{k})=m_{0}-\sum_{i}\beta_{i}\cos k_{i}$
and $\tau_{i}$ and $\sigma_{i}$ are Pauli matrices acting on orbital
and spin space, respectively. $H(\boldsymbol{k})$ preserves time-reversal
($\mathcal{T}=i\sigma_{y}\mathbb{K}$), reflection about the $xz$
and $yz$ planes ($M_{i\to-i}=\tau_{z}\sigma_{i}$, $i=x,y,z$) and
twofold rotation about the $z$-axis ($R_{i}=\sigma_{i}$), but breaks
inversion ($\mathcal{I}=\tau_{z}$), reflection about the $xy$ plane,
and twofold rotation about the $x$ and the $y$ axes are broken.
Its spectrum is given by 
\begin{equation}
\left(E(\boldsymbol{k})+\mu\right)^{2}=v_{z}^{2}\sin k_{z}^{2}+m^{2}(\boldsymbol{k})+\left(\sqrt{v_{x}^{2}\sin k_{x}^{2}+v_{y}^{2}\sin k_{y}^{2}}\pm\ell\right)^{2}\label{eq:normal spectrum-1}
\end{equation}
Defining $X=\cos k_{x}$, $Y=\cos k_{y}$, a quadruplet of Weyl nodes
(WNs) appears in the $k_{z}=0$ or $\pi$ plane at $(K_{x},K_{y})=(\pm\cos^{-1}X,\pm\cos^{-1}Y)$
for each intersection between the following ellipse and lines within
the unit square $X\in[-1,1]$, $Y\in[-1,1]$ 
\begin{align}
v_{x}^{2}X^{2}+v_{y}^{2}Y^{2} & =v_{x}^{2}+v_{y}^{2}-\ell^{2}\label{eq:Ellipse-1}\\
\beta_{x}X+\beta_{y}Y & =M_{k_{z}}=m_{0}-\beta_{z}\cos k_{z}\label{eq:Lines-1}
\end{align}
When the ellipse and line do not intersect within the unit square,
the system is an $\mathcal{T}$-symmetric insulator. These behaviors
are depicted in the top panel of Fig. \ref{fig:Ellipse-Line-FermiArc-1}

At $\ell=0$, $\mathcal{I}$ is restored, the system is necessarily
insulating since the ellipse circumscribes the unit square and the
topological nature of the insulator can be deduced from the parity
criterion which only depends on $\text{sgn}[m(\boldsymbol{k})]$ at
the eight time-reversal invariant momenta $(0/\pi,0/\pi,0/\pi)$.
For larger $\ell$, the strong topological index of an insulating
state can be obtained easily by observing the connectivity of the
Fermi arcs on an $xy$-surface, as shown in the bottom panel of Fig.
\ref{fig:Ellipse-Line-FermiArc-1}. Imagine tuning a parameter that
creates and subsequently annihilates a quadruplet of WNs. Now, nodes
are always created as well as annihilated in pairs of opposite chirality.
Moreover, creating a pair of nodes and moving them apart leaves behind
a surface Fermi arc that connects the surface projections of the nodes.
If the nodes switch partners between creation and annihilation --
in other words, if a given right-handed WN is created along with a
left-handed WN but annihilates a different left-handed WN -- a non-degenerate,
$\mathcal{T}$-invariant Fermi surface is left behind on the surface.
Such a Fermi surface can be viewed as the surface state of a topological
insulator doped away from charge neutrality. Therefore, each time
WNs switch partners between creation and annihilation, the strong
topological index of the bulk insulator toggles.

In the main paper, we choose parameters such that the line defined
by $m(k_{z}=\pi)=0$ never intersects the ellipse. Then, all the normal
state phase transitions occur via crossings in the $k_{z}=0$ plane,
which gives access to trivial and topological insulators as well as
T-WSMs with $N=1,2$.

\section{\label{B}Vortex topological invariant in a minimal lattice model\label{sec:Topological-vortex-N}}

In this section, we use a perturbative scheme to explicitly determine
the topological state of the vortex in the lattice model (\ref{eq:lattice model-1})
in the range of parameters which gives $N=1$ quadruplet of WNs.

\subsection{Reduction to a canonical Weyl Hamiltonian\label{B1}}

We begin with the Bloch Hamiltonian (\ref{eq:lattice model-1}) and
assume the parameters are chosen so that there is a single quadruplet
of WNs, at $(\pm K_{x},\pm K_{y},0)$. The Bloch Hamiltonian at these
points has a higher symmetry, namely, $[H(\boldsymbol{K}),\tau_{y}\sigma_{z}]=0$,
so it is convenient to work in the eigenbasis of $\tau_{y}\sigma_{z}$.
For convenience, let us perform a rotation 
\begin{equation}
H'(\boldsymbol{K})=e^{i\tau_{x}\pi/4}H(\boldsymbol{K})e^{-i\tau_{x}\pi/4}=\tau_{x}\boldsymbol{\sigma}\cdot\boldsymbol{d}(\boldsymbol{k})+\tau_{z}\sigma_{z}\ell-\mu
\end{equation}
which explicitly diagonalizes the term proportional to $\ell$. Since
$\left|\boldsymbol{d}(\boldsymbol{K})\right|^{2}=\ell^{2}$ at the
nodes according to Equation (4) of the main paper, the four states at each
WN have energies $2\ell,0,0,-2\ell$. The two zero energy states explicitly
are 
\begin{equation}
|A'\rangle=\frac{1}{\sqrt{2}}\left(1,0,0,-e^{i\theta_{d}}\right)^{T}\,\,,\,\,|B'\rangle=\frac{1}{\sqrt{2}}\left(0,e^{i\theta_{d}},1,0\right)^{T}
\end{equation}
where $\theta_{d}=\arg(d_{x}+id_{y})$ and the primes serve as reminders
that we have performed a $e^{i\tau_{x}\pi/4}$ rotation. The low energy
Hamiltonian near the WN in the $(|A'\rangle,|B'\rangle)^{T}$ basis
is given by 
\begin{equation}
H_{W}^{\prime}(\boldsymbol{p})=\left(\Sigma_{x},\Sigma_{y},\Sigma_{z}\right)\left(\begin{array}{ccc}
0 & 0 & v_{z}\\
\beta_{x}\sin K_{x} & \beta_{y}\sin K_{y} & 0\\
-v_{x}^{2}\ell^{-1}\sin K_{x}\cos K_{x} & -v_{y}^{2}\ell^{-1}\sin K_{y}\cos K_{y} & 0
\end{array}\right)\left(\begin{array}{c}
p_{x}\\
p_{y}\\
p_{z}
\end{array}\right)-\mu
\end{equation}
where $\Sigma_{i}$ are Pauli operators in the $|A'\rangle,|B'\rangle$
basis. Note that reversing $K_{J}$ to get to a different WN is equivalent
to reversing $p_{J}$ in $H_{W}^{\prime}$. At $p_{z}=0$, $H_{W}^{\prime}$
contains only $\Sigma_{z}$ and $\Sigma_{y}$. For convenience, we
rotate $\Sigma_{z}\to-\Sigma_{x}$, $\Sigma_{x}\to\Sigma_{z}$ to
define $H_{W}^{\prime\prime}=e^{i\Sigma_{y}\pi/4}H_{W}^{\prime}e^{-i\Sigma_{y}\pi/4}$.
$H_{W}^{\prime\prime}$ in the $p_{z}=0$ plane is 
\begin{equation}
H_{W}^{\prime\prime}(p_{z}=0)=\ell^{-1}\left(\Sigma_{x},\Sigma_{y}\right)\hat{M}\left(\begin{array}{c}
p_{x}\\
p_{y}
\end{array}\right)-\mu\label{eq:non-canonical}
\end{equation}
where $\hat{M}=\left(\begin{array}{cc}
v_{x}^{2}\ell^{-1}\sin K_{x}\cos K_{x} & v_{y}^{2}\ell^{-1}\sin K_{y}\cos K_{y}\\
\beta_{x}\sin K_{x} & \beta_{y}\sin K_{y}
\end{array}\right)$. To bring this into a canonical form, we perform a singular value
decomposition of $\hat{M}$ 
\begin{equation}
\hat{M}=R(\phi_{\Sigma})\left(\begin{array}{cc}
v_{X} & 0\\
0 & v_{Y}
\end{array}\right)R^{T}(\phi_{p})
\end{equation}
where $R(\phi)=\left(\begin{array}{cc}
\cos\phi & -\sin\phi\\
\sin\phi & \cos\phi
\end{array}\right)$ and $v_{X,Y}>0$. We have assumed that the WNs at $\pm(K_{x},K_{y})$
can be brought into a canonical form by proper rotations. This automatically
means that the nodes at $\pm(K_{x},-K_{y})$ need improper rotations.%
{} The necessity of singular value decomposition indicates that the
principal axes for $\boldsymbol{p}$ and $\boldsymbol{\Sigma}$ are
different, and both differ from the Cartesian axes of the original
problem. Moreover, $v_{X}\neq v_{Y}$, implying that the WN is anisotropic.
Nonetheless, this can be brought into a canonical form $H_{W}^{\prime\prime}=v_{X}\Sigma_{X}P_{X}+v_{Y}\Sigma_{Y}P_{Y}-\mu$
through the rotations

\begin{align}
\left(\begin{array}{c}
P_{X}\\
P_{Y}
\end{array}\right) & =R^{T}(\phi_{p})\left(\begin{array}{c}
p_{x}\\
p_{y}
\end{array}\right)\\
\left(\begin{array}{c}
X\\
Y
\end{array}\right) & =R^{T}(\phi_{p})\left(\begin{array}{c}
x\\
y
\end{array}\right)\\
\left(\begin{array}{c}
\Sigma_{X}\\
\Sigma_{Y}
\end{array}\right) & =R^{T}(\phi_{\Sigma})\left(\begin{array}{c}
\Sigma_{x}\\
\Sigma_{y}
\end{array}\right)=e^{-i\Sigma_{z}\phi_{\Sigma}/2}\left(\begin{array}{c}
\Sigma_{x}\\
\Sigma_{y}
\end{array}\right)e^{i\Sigma_{z}\phi_{\Sigma}/2}
\end{align}

\subsection{Vortex modes of anisotropic vortex \label{subsec:Vortex-modes-of-anisotropic-vortex}\label{B2}}

In the presence of $s$-wave superconductivity, the Bogoliubov-deGennes
Hamiltonian is given by 
\begin{equation}
H_{BdG}^{\prime\prime}(\boldsymbol{P})=\left(\begin{array}{cc}
H_{W}^{\prime\prime}(\boldsymbol{P}) & \Delta(\boldsymbol{R})\\
\Delta^{*}(\boldsymbol{R}) & -H_{W}^{\prime\prime}(\boldsymbol{P})
\end{array}\right)
\end{equation}
in the basis $\frac{1}{\sqrt{2}}\left(c_{A'}+c_{B'},-c_{A'}+c_{B'},-c_{A'}^{\dagger}+c_{B'}^{\dagger},-c_{A'}^{\dagger}-c_{B'}^{\dagger}\right)^{T}$.
Furthermore, if the superconductivity develops a vortex $\Delta(\boldsymbol{r})=\Delta_{0}(r)e^{i\theta}$,
where $\theta=\arg(x+iy)$, the pairing term in $H_{BdG}^{\prime\prime}$
becomes $\Delta(\boldsymbol{R})=e^{i(\phi_{\Sigma}+\Theta)}\Delta_{0}(R)$,
where $\Theta=\arg(X+iY)$. If $v_{X}=v_{Y}$, the problem has a rotational
symmetry which can be used to obtain the eigenmodes of $H_{BdG}^{\prime\prime}$
analytically. This result is well-known \cite{Fu2008,Hosur2011,Read2000}.
When $v_{X}\neq v_{Y}$, we can still obtain the eigenmodes analytically
in the linear approximation $\Delta_{0}(R)=\Delta_{0}R/\xi$, where
$\xi$ is the superconducting coherence length.

We explicitly write 
\begin{equation}
H_{BdG}^{\prime\prime}(\boldsymbol{P})=\Pi_{z}H_{W}^{\prime\prime}(\boldsymbol{P})+\frac{\Delta_{0}R}{\xi}\left(\Pi_{x}\cos\left(\Theta+\phi_{\Sigma}\right)-\Pi_{y}\sin\left(\Theta+\phi_{\Sigma}\right)\right)
\end{equation}
The $\phi_{\Sigma}$-dependence can be eliminated by a $\Pi_{z}$-rotation:
\begin{align}
H_{BdG}^{\prime\prime\prime}(\boldsymbol{P}) & =e^{-i\Pi_{z}\phi_{\Sigma}/2}H^{\prime\prime}(\boldsymbol{P})e^{i\Pi_{z}\phi_{\Sigma}/2}\\
 & =\Pi_{z}H_{W}^{\prime\prime}(\boldsymbol{P})+\frac{\Delta_{0}}{\xi}\left(\Pi_{x}X-\Pi_{y}Y\right)
\end{align}
At $\mu=0$, we can separate the $X$ and $Y$ parts of the problem
via another rotation. Specifically, define 
\begin{align}
H_{BdG}^{\prime\prime\prime\prime}(\boldsymbol{P}) & =e^{i\Pi_{y}\Sigma_{Y}\pi/4}H_{BdG}^{\prime\prime\prime}\left(\boldsymbol{P}\right)e^{-i\Pi_{y}\Sigma_{Y}\pi/4}\\
 & =\Pi_{z}\left(v_{X}\Sigma_{X}P_{X}+\frac{\Delta_{0}}{\xi}\Sigma_{Y}X\right)-\left(\Pi_{x}v_{Y}P_{Y}+\Pi_{y}\frac{\Delta_{0}}{\xi}Y\right)\\
 & =\sqrt{\frac{2\Delta_{0}}{\xi}}\left(\begin{array}{cccc}
 & -i\sqrt{v_{X}}a_{X} & i\sqrt{v_{Y}}a_{Y}\\
i\sqrt{v_{X}}a_{X}^{\dagger} &  &  & i\sqrt{v_{Y}}a_{Y}\\
-i\sqrt{v_{Y}}a_{Y}^{\dagger} &  &  & i\sqrt{v_{X}}a_{X}\\
 & -i\sqrt{v_{Y}}a_{Y}^{\dagger} & -i\sqrt{v_{X}}a_{X}^{\dagger}
\end{array}\right)
\end{align}
where $a_{J}=\sqrt{\frac{\xi}{2\Delta_{0}v_{J}}}\left(\frac{\Delta_{0}}{\xi}J+iv_{J}P_{J}\right),J=X,Y$
is the usual annihilation operator for a quantum harmonic oscillator.
The eigenstates of $H_{BdG}^{\prime\prime\prime\prime}$ are of the
form $\left(|n_{X}-1,n_{Y}-1\rangle,|n_{X},n_{Y}-1\rangle,|n_{X}-1,n_{Y}\rangle,|n_{X},n_{Y}\rangle\right)^{T}$.
In this basis, 
\begin{align}
H_{BdG}^{\prime\prime\prime\prime}(n_{X},n_{Y}) & =\sqrt{\frac{2\Delta_{0}}{\xi}}\left(\begin{array}{cccc}
 & -i\sqrt{v_{X}n_{X}} & i\sqrt{v_{Y}n_{Y}}\\
i\sqrt{v_{X}n_{X}} &  &  & i\sqrt{v_{Y}n_{Y}}\\
-i\sqrt{v_{Y}n_{Y}} &  &  & i\sqrt{v_{X}n_{X}}\\
 & -i\sqrt{v_{Y}n_{Y}} & -i\sqrt{v_{X}n_{X}}
\end{array}\right)\\
 & =\sqrt{\frac{2\Delta_{0}}{\xi}}\left(\Pi_{z}\Sigma_{Y}\sqrt{v_{X}n_{X}}+\Pi_{x}\sqrt{v_{Y}n_{Y}}\right)
\end{align}
Thus, it has the spectrum 
\begin{equation}
E\left(n_{X},n_{Y}\right)=\pm\sqrt{\frac{2\Delta_{0}}{\xi}\left(v_{X}n_{X}+v_{Y}n_{Y}\right)}\label{eq:gap}
\end{equation}

In particular, the zero mode is given by $n_{X}=n_{Y}=0$ and has
the wavefunction 
\begin{equation}
\varphi^{\prime\prime\prime\prime}(\boldsymbol{R})=(0,0,0,|0,0\rangle)^{T}\equiv(0,0,0,1)^{T}f_{00}(X,Y)
\end{equation}
where $f_{00}\left(X,Y\right)=\sqrt{\dfrac{\Delta_{0}}{\pi\xi\sqrt{v_{x}v_{y}}}}\exp\left[-\dfrac{\Delta_{0}}{2\xi}\left(\dfrac{X^{2}}{v_{X}}+\dfrac{Y^{2}}{v_{Y}}\right)\right]$
is the wavefunction for the $(n_{X}=0,n_{Y}=0)$ mode of the 2D harmonic
oscillator. Undoing the rotations generated by $\Pi_{y}\Sigma_{Y}$,
$\Pi_{z}$, $\Sigma_{y}$ and the singular value decomposition gives
\begin{align}
\varphi^{\prime}(\boldsymbol{R}) & =e^{-i\Sigma_{y}\pi/4}e^{i\Pi_{z}\Sigma_{z}\phi_{\Sigma}/2}\varphi^{\prime\prime}=\frac{1}{\sqrt{2}}e^{-i\Sigma_{y}\pi/4}\left(ie^{i\phi_{\Sigma}},0,0,1\right)^{T}f_{00}(X,Y)\\
\varphi^{\prime}(x,y) & =\frac{1}{2}\left(ie^{i\phi_{\Sigma}},ie^{i\phi_{\Sigma}},-1,1\right)^{T}\tilde{f}(x,y)
\end{align}
in the basis $(c_{A},c_{B},c_{B}^{\dagger},-c_{A}^{\dagger})$, where
$\tilde{f}(x,y)=\sqrt{\frac{\Delta_{0}}{\pi\xi\sqrt{v_{X}v_{Y}}}}\exp\left[-\frac{\Delta_{0}r^{2}}{4\xi}\left\{ \left(\frac{1}{v_{X}}+\frac{1}{v_{Y}}\right)+\left(\frac{1}{v_{X}}-\frac{1}{v_{Y}}\right)\cos[2(\theta+\phi_{p})]\right\} \right]\equiv\sqrt{\frac{\Delta_{0}}{\pi\xi\sqrt{v_{X}v_{Y}}}}\exp\left[-\frac{\Delta_{0}r^{2}}{4\xi}\left\{ \frac{1}{v_{+}}+\frac{1}{v_{-}}\cos[2(\theta+\phi_{p})]\right\} \right]$.
$\varphi^{\prime}$ is an eigenstate of charge conjugation: $\mathbb{C}\varphi^{\prime}\equiv\Pi_{y}\Sigma_{y}\varphi^{\prime*}=ie^{-i\phi_{\Sigma}}\varphi^{\prime}$
and hence, represents a Majorana mode. In the original basis $\left(c_{s\uparrow},c_{s_{\downarrow}},c_{p\uparrow},c_{p\downarrow},c_{s\downarrow}^{\dagger},-c_{s_{\uparrow}}^{\dagger},c_{p\downarrow}^{\dagger},-c_{p\uparrow}^{\dagger}\right)^{T}$,

\begin{align}
\varphi(x,y) & =\frac{e^{-i\pi/4}}{2\sqrt{2}}\left(-e^{i\phi_{\Sigma}},ie^{i(\theta_{d}+\phi_{\Sigma})},-e^{i\phi_{\Sigma}},-ie^{i(\theta_{d}+\phi_{\Sigma})},-ie^{-i\theta_{d}},1,ie^{-i\theta_{d}},1\right)^{T}\tilde{f}(x,y)\\
 & \equiv\chi\tilde{f}(x,y)\nonumber 
\end{align}
Finally, $\chi^{\dagger}\Pi_{z}\tau_{x}\sigma_{z}\chi=1$, so non-zero
$k_{z}$ induces a dispersion $E(k_{z})=v_{z}\sin k_{z}$ and thus
produces a CMM.

\subsection{Hybridization between CMMs\label{B3}}

The Majorana fermions coming from WNs at $(K_{x},-K_{y})$, $(-K_{x},-K_{y})$
and $(-K_{x},K_{y})$ can be obtained by applying the symmetry operations
$S_{y}=i\mathcal{T}M_{y}=\tau_{z}\mathbb{K}\otimes y\to-y$, $S_{xy}=-i\Pi_{z}M_{x}M_{y}=\Pi_{z}\sigma_{z}\otimes(x,y)\to-(x,y)$
and $S_{x}=\Pi_{z}\mathcal{T}M_{x}=\Pi_{z}\tau_{z}\sigma_{z}\mathbb{K}\otimes x\to-x$,
respectively. The result after reinstating the fast spatial variation
is 
\begin{align*}
\psi_{\lambda_{x}\lambda_{y}}(x,y) & =\frac{1}{2\sqrt{2}}e^{i(\lambda_{x}K_{x}x+\lambda_{y}K_{y}y-\lambda_{x}\lambda_{y}\pi/4)}\sqrt{\frac{\Delta_{0}}{\pi\xi\sqrt{v_{X}v_{Y}}}}\exp\left[-\frac{\Delta_{0}r^{2}}{4\xi}\left\{ \frac{1}{v_{+}}+\frac{\cos[2(\theta-\lambda_{x}\lambda_{y}\phi_{p})]}{v_{-}}\right\} \right]\times\\
 & \left(-e^{i\lambda_{x}\lambda_{y}\phi_{\Sigma}},i\lambda_{y}e^{i\lambda_{x}\lambda_{y}(\theta_{d}+\phi_{\Sigma})},-\lambda_{x}\lambda_{y}e^{i\lambda_{x}\lambda_{y}\phi_{\Sigma}},-i\lambda_{x}e^{i\lambda_{x}\lambda_{y}(\theta_{d}+\phi_{\Sigma})},-i\lambda_{y}e^{-i\lambda_{x}\lambda_{y}\theta_{d}},1,i\lambda_{x}e^{-i\lambda_{x}\lambda_{y}\theta_{d}},\lambda_{x}\lambda_{y}\right)^{T}
\end{align*}
where $\lambda_{x}=\pm$, $\lambda_{y}=\pm$. For $K_{x,y}\xi\gg1$,
the leading perturbation from band curvature is given by the matrix
elements 
\begin{align}
\left\langle \psi_{\lambda_{x}^{\prime}\lambda_{y}^{\prime}}\left|H_{2}\right|\psi_{\lambda_{x}\lambda_{y}}\right\rangle  & =e^{i\pi/4(\lambda_{x}^{\prime}\lambda_{y}^{\prime}-\lambda_{x}\lambda_{y})}\intop_{x,y}e^{i[(\lambda_{x}-\lambda_{x}^{\prime})K_{x}x+(\lambda_{y}-\lambda_{y}^{\prime})K_{y}y]}\varphi_{\lambda_{x}^{\prime}\lambda_{y}^{\prime}}^{\dagger}(x,y)\times\nonumber \\
 & \,\,\,\,\,\dfrac{1}{2}\left(p_{x}^{2}\partial_{k_{x}}^{2}H_{BdG}(\lambda_{x}K_{x},\lambda_{y}K_{y})+p_{y}^{2}\partial_{k_{y}}^{2}H_{BdG}(\lambda_{x}K_{x},\lambda_{y}K_{y})\right)\varphi_{\lambda_{x}\lambda_{y}}(x,y)\\
 & =-\dfrac{1}{2}e^{i\pi/4(\lambda_{x}^{\prime}\lambda_{y}^{\prime}-\lambda_{x}\lambda_{y})}\intop_{x,y}e^{i[(\lambda_{x}-\lambda_{x}^{\prime})K_{x}x+(\lambda_{y}-\lambda_{y}^{\prime})K_{y}y]}\varphi_{\lambda_{x}^{\prime}\lambda_{y}^{\prime}}^{\dagger}(x,y)\Pi_{z}\times\nonumber \\
 & \,\,\left[\begin{array}{l}
p_{x}^{2}\left(\frac{\lambda_{x}+\lambda_{x}^{\prime}}{2}\tau_{x}\sigma_{x}v_{x}\sin K_{x}-\tau_{z}\beta_{x}\cos K_{x}\right)+\\
\ \ \ \ \ \ \ \ \ p_{y}^{2}\left(\frac{\lambda_{y}+\lambda_{y}^{\prime}}{2}\tau_{x}\sigma_{y}v_{y}\sin K_{y}-\tau_{z}\beta_{y}\cos K_{y}\right)
\end{array}\right]\varphi_{\lambda_{x}\lambda_{y}}(x,y)
\end{align}
which consists of straightforward Gaussian integrals. First, let us
compute the spinor products. In the basis $\left(|\chi_{++}\rangle,|\chi_{--}\rangle,|\chi_{-+}\rangle,|\chi_{+-}\rangle\right)$,
we find 
\begin{align}
(\lambda_{i}+\lambda_{i^{\prime}})\Pi_{z}\tau_{x}\sigma_{i} & \to0;i=x,y
\end{align}
\begin{eqnarray}
\Pi_{z}\tau_{z} & \to & \left(\begin{array}{cccc}
 &  & e^{-i\phi_{\Sigma}}\cos\theta_{d}\sin(\phi_{\Sigma}+\theta_{d}) & -e^{-i\phi_{\Sigma}}\cos(\phi_{\Sigma}+\theta_{d})\sin\theta_{d}\\
 &  & -e^{-i\phi_{\Sigma}}\cos(\phi_{\Sigma}+\theta_{d})\sin\theta_{d} & e^{-i\phi_{\Sigma}}\cos\theta_{d}\sin(\phi_{\Sigma}+\theta_{d})\\
 & 0\\
0
\end{array}\right)+\\
 &  & \ \ \ \ \left(\begin{array}{cccc}
 &  &  & 0\\
 &  & 0\\
e^{i\phi_{\Sigma}}\cos\theta_{d}\sin(\phi_{\Sigma}+\theta_{d}) & -e^{i\phi_{\Sigma}}\cos(\phi_{\Sigma}+\theta_{d})\sin\theta_{d}\\
-e^{i\phi_{\Sigma}}\cos(\phi_{\Sigma}+\theta_{d})\sin\theta_{d} & e^{i\phi_{\Sigma}}\cos\theta_{d}\sin(\phi_{\Sigma}+\theta_{d})
\end{array}\right)\nonumber 
\end{eqnarray}

Note that only CMMs coming from nodes of opposite chiralities mix,
in which case $\lambda_{x}^{\prime}\lambda_{y}^{\prime}=-\lambda_{x}\lambda_{y}$,
and the hybridization is caused by the ``mass'' term, not the ``kinetic''
terms, of the Dirac Hamiltonian (\ref{eq:lattice model-1}). Finally,
we get the effective Hamiltonian in the basis $\left(|\psi_{++}\rangle,|\psi_{--}\rangle,e^{-i\phi_{\Sigma}}|\psi_{-+}\rangle,e^{-i\phi_{\Sigma}}|\psi_{+-}\rangle\right)^{T}$
as 
\begin{equation}
H_{eff}=i\left(\begin{array}{cccc}
 &  & q_{x} & q_{y}\\
 &  & q_{y} & q_{x}\\
-q_{x} & -q_{y}\\
-q_{y} & -q_{x}
\end{array}\right)\label{eq:H_effective}
\end{equation}
where $q_{i}$ decays as a Gaussian as a function of $\Delta K_{i}=K_{i}-(-K_{i})$.
Explicitly, 
\begin{align}
q_{x} & =\frac{\left(\Delta K_{x}\right)^{2}\cos\theta_{d}\sin\left(\phi_{\Sigma}+\theta_{d}\right)}{2}\exp\left[-\frac{\xi\left(\Delta K_{x}\right)^{2}}{2\Delta_{0}\left(\frac{1}{v_{+}}+\frac{\cos2\phi_{p}}{v_{-}}\right)}\right]\times\nonumber \\
 & \ \ \ \ \!\ \left(\mathbb{V}_{+}^{2}\left(\frac{1}{v_{+}}-\frac{\cos2\phi_{p}}{v_{-}}\right)^{2}\beta_{x}\cos K_{x}+\mathbb{V}_{-}^{2}\frac{\sin^{2}2\phi_{p}}{v_{-}^{2}}\beta_{y}\cos K_{y}\right)\\
q_{y} & =-\frac{\left(\Delta K_{y}\right)^{2}\cos\left(\phi_{\Sigma}+\theta_{d}\right)\sin\theta_{d}}{2}\exp\left[-\frac{\xi\left(\Delta K_{y}\right)^{2}}{2\Delta_{0}\left(\frac{1}{v_{+}}-\frac{\cos2\phi_{p}}{v_{-}}\right)}\right]\times\nonumber \\
 & \ \!\ \ \!\ \left(\mathbb{V}_{+}^{2}\frac{\sin^{2}2\phi_{p}}{v_{-}^{2}}\beta_{x}\cos K_{x}+\mathbb{V}_{-}^{2}\left(\frac{1}{v_{+}}+\frac{\cos2\phi_{p}}{v_{-}}\right)^{2}\beta_{y}\cos K_{y}\right)\label{eq:qy}
\end{align}
to leading order in $\Delta_{0}\xi/(\Delta K_{x})^{2}$ and $\Delta_{0}\xi/(\Delta K_{y})^{2}$,
where $\mathbb{V}_{\pm}^{2}=(v_{X}v_{Y})^{-1/2}\left(\frac{1}{v_{+}}\pm\frac{\cos2\phi_{p}}{v_{-}}\right)^{-1/2}\left(\frac{1}{v_{+}}\mp\frac{\cos2\phi_{p}}{v_{-}}\right)^{-5/2}$
has units of velocity-squared. For $|\Delta K_{x}|\gg|\Delta K_{y}|$
and $|\Delta K_{x}|\ll|\Delta K_{y}|$, $Q=\left(\begin{array}{cc}
q_{x} & q_{y}\\
q_{y} & q_{x}
\end{array}\right)$ simplifies to $\left(\begin{array}{cc}
0 & q_{y}\\
q_{y} & 0
\end{array}\right)$ and $\left(\begin{array}{cc}
q_{x} & 0\\
0 & q_{x}
\end{array}\right)$, respectively, so that the vortex topological invariant is 
\begin{equation}
\nu=\text{sgn}[\det(Q)]=\begin{cases}
-1 & |\Delta K_{x}|\gg|\Delta K_{y}|\\
+1 & |\Delta K_{x}|\ll|\Delta K_{y}|
\end{cases}
\end{equation}
Thus, there is vortex phase transition as the WNs ``switch partners'',
i.e., the nearest WN to a given WN changes.

\subsection{Hybridization gap in the continuum limit\label{B4}}

Deep in the topological phase, $q_{x}\ll q_{y}$, and the eigenvalues
of $H_{eff}$ reduce to $\pm|q_{y}|$. To detect the surface Majorana
fermion in this regime, one must thus be at temperatures $T\ll|q_{y}|/k_{B}$,
which we now estimate.

$|q_{y}|$, as given by (\ref{eq:qy}), is straightforward albeit
tedious to estimate in the continuum limit $k_{x},k_{y}\ll1$. Restricting
to the $k_{z}=0$ plane as done throughout the derivations above,
{} (\ref{eq:Ellipse-1}) and (\ref{eq:Lines-1}), which determine the
locations of the WNs, simplify to 
\begin{align}
v_{x}^{2}K_{x}^{2}+v_{y}^{2}K_{y}^{2} & =\ell^{2}\label{eq:Ellipse-1-1}\\
\beta_{x}K_{x}^{2}+\beta_{y}K_{y}^{2} & =-2m_{\Gamma}\label{eq:Lines-1-1}
\end{align}
where $m_{\Gamma}=m_{0}-\sum_{i=x,y,z}\beta_{i}$. Then, $\hat{M}$,
which defines the non-canonical Weyl Hamiltonian $H_{W}^{\prime\prime}$
in (\ref{eq:non-canonical}) becomes
\begin{equation}
\hat{M}=\left(\begin{array}{cc}
v_{x}^{2}K_{x}/\ell & v_{y}^{2}K_{y}/\ell\\
\beta_{x}K_{x} & \beta_{y}K_{y}
\end{array}\right)=R(\phi_{\Sigma})\left(\begin{array}{cc}
v_{X} & 0\\
0 & v_{Y}
\end{array}\right)R^{T}(\phi_{p})
\end{equation}
Deep in the topological phase of the vortex where $K_{y}^{2}\ll K_{x}^{2}$,
(\ref{eq:Ellipse-1-1}) implies $\ell\approx K_{x}v_{x}$. To ensure
consistency with (\ref{eq:Lines-1-1}), we must fine-tune $m_{\Gamma}\approx-\beta_{x}K_{x}^{2}/2$.
Note that $|\ell|,|m_{\Gamma}|\ll|v_{x}|,|v_{y}|$ ensures that the
WNs occur in the continuum limit, $K_{x,y}\ll1$. Within the continuum
limit, $K_{y}\ll K_{x}$ is enforced by fine-tuning $\beta_{x}$ without
making any of the individual parameters -- $\ell,$ $v_{x}$, $v_{y}$,
$\beta_{x}$, $\beta_{y}$ and $m_{\Gamma}$ -- further small or
large in magnitude. In this regime, $\hat{M}$ simplifies to
\begin{equation}
\hat{M}\approx\left(\begin{array}{cc}
v_{x} & v_{y}^{2}K_{y}/v_{x}K_{x}\\
\beta_{x}K_{x} & \beta_{y}K_{y}
\end{array}\right)=R(\phi_{\Sigma})\left(\begin{array}{cc}
v_{X} & 0\\
0 & v_{Y}
\end{array}\right)R^{T}(\phi_{p})
\end{equation}
where $v_{X},v_{Y}>0$, assuming that the singular value decomposition
at $(K_{x},K_{y})$ is achieved by proper rotations.%
{} Explicitly,
\begin{align}
v_{X} & \approx\left|v_{x}^{2}+\beta_{x}^{2}K_{x}^{2}+\frac{(v_{y}^{2}/K_{x}+\beta_{x}\beta_{y}K_{x})^{2}}{v_{x}^{2}+\beta_{x}^{2}K_{x}^{2}}K_{y}^{2}\right|^{1/2}\approx v_{x}\\
v_{Y} & \approx\frac{|v_{y}^{2}\beta_{x}/v_{x}-v_{x}\beta_{y}|}{v_{x}^{2}+\beta_{x}^{2}K_{x}^{2}}|K_{y}|\approx\left|\beta_{x}\frac{v_{y}^{2}}{v_{x}^{2}}-\beta_{y}\right||K_{y}|\\
|\phi_{p}| & \approx\left|\frac{v_{y}^{2}+\beta_{x}\beta_{y}K_{x}^{2}}{v_{x}^{2}+\beta_{x}^{2}K_{x}^{2}}\frac{K_{y}}{K_{x}}\right|\approx\left|\frac{v_{y}^{2}K_{y}}{v_{x}^{2}K_{x}}\right|\\
|\phi_{\Sigma}| & \approx\left|\frac{\beta_{x}^{2}K_{x}^{2}+\beta_{y}^{2}K_{y}^{2}}{v_{x}^{2}+v_{y}^{4}K_{y}^{2}/v_{x}^{2}K_{x}^{2}}\right|^{1/2}\approx\left|\frac{\beta_{x}K_{x}}{v_{x}}\right|
\end{align}
Since $v_{Y}/v_{X},|\phi_{p}|\sim O(K_{y}/K_{x})\ll1$, this further
yields 
\begin{align}
\frac{1}{v_{\pm}} & \approx\pm\frac{1}{v_{Y}}\\
\mathbb{V}_{+} & \approx\frac{v_{Y}}{2\sqrt{2}}\\
\mathbb{V}_{-} & \approx\frac{v_{X}}{2\sqrt{2}}\\
\frac{\mathbb{V}_{+}}{\mathbb{V}_{-}} & \approx\frac{v_{Y}}{v_{X}}\sim O\left(\frac{K_{y}}{K_{x}}\right)\ll1
\end{align}
Putting all this together along with $|\phi_{\Sigma}|\sim O(K_{x})\ll1$
in the continuum limit and $|\theta_{d}|\approx|v_{y}K_{y}/v_{x}K_{x}|\ll1$
deep in the topological phase, we get for $|q_{y}|$:
\begin{equation}
|q_{y}|=\dfrac{1}{4}\beta_{y}\left(\Delta K_{y}\right)^{2}|\theta_{d}|\exp\left[-\frac{\xi v_{Y}\left(\Delta K_{y}\right)^{2}}{4\Delta_{0}}\right]
\end{equation}
{} Assuming typical values for band parameters ($v_{Y}\sim v_{X}/10\sim10^{4}m/s$,
$K_{y}\sim K_{x}/10\sim0.02\times2\pi/a$ with $a\approx6.0\text{A}$,
$v_{y}=v_{x}$ so that $|\theta_{d}|\sim0.1$ and $\beta_{y}^{-1}\sim$bare
electron mass) and superconducting properties ($\Delta_{0}\sim5K$
and $\xi\sim5nm$) gives $|q_{y}|\sim0.1K$ and $|q_{x}|\sim0$ with
$|\Delta K_{y}\xi|\sim2\gtrsim1$. The vortex originating from individual
WNs is $\delta\sim\Delta_{0}\hbar v_{Y}/\mu\xi\sim1K$ for $\mu\sim100K$,
so that $|q_{y}|\lesssim\delta\lesssim\Delta_{0}$, ensuring that
we are in the right perturbative regimes. Note that this expression
for $\delta$ differs from the one that follows from (\ref{eq:gap});
the latter is only valid at $\mu=0$ whereas real materials invariable
have non-zero $\mu\gg\Delta_{0}$ and fall in the regime where the
former is valid.

\section{Level crossings, Pfaffian and Topological phase transitions\label{C}}

\begin{figure}[h]
\subfloat[]{\includegraphics[width=0.4\linewidth]{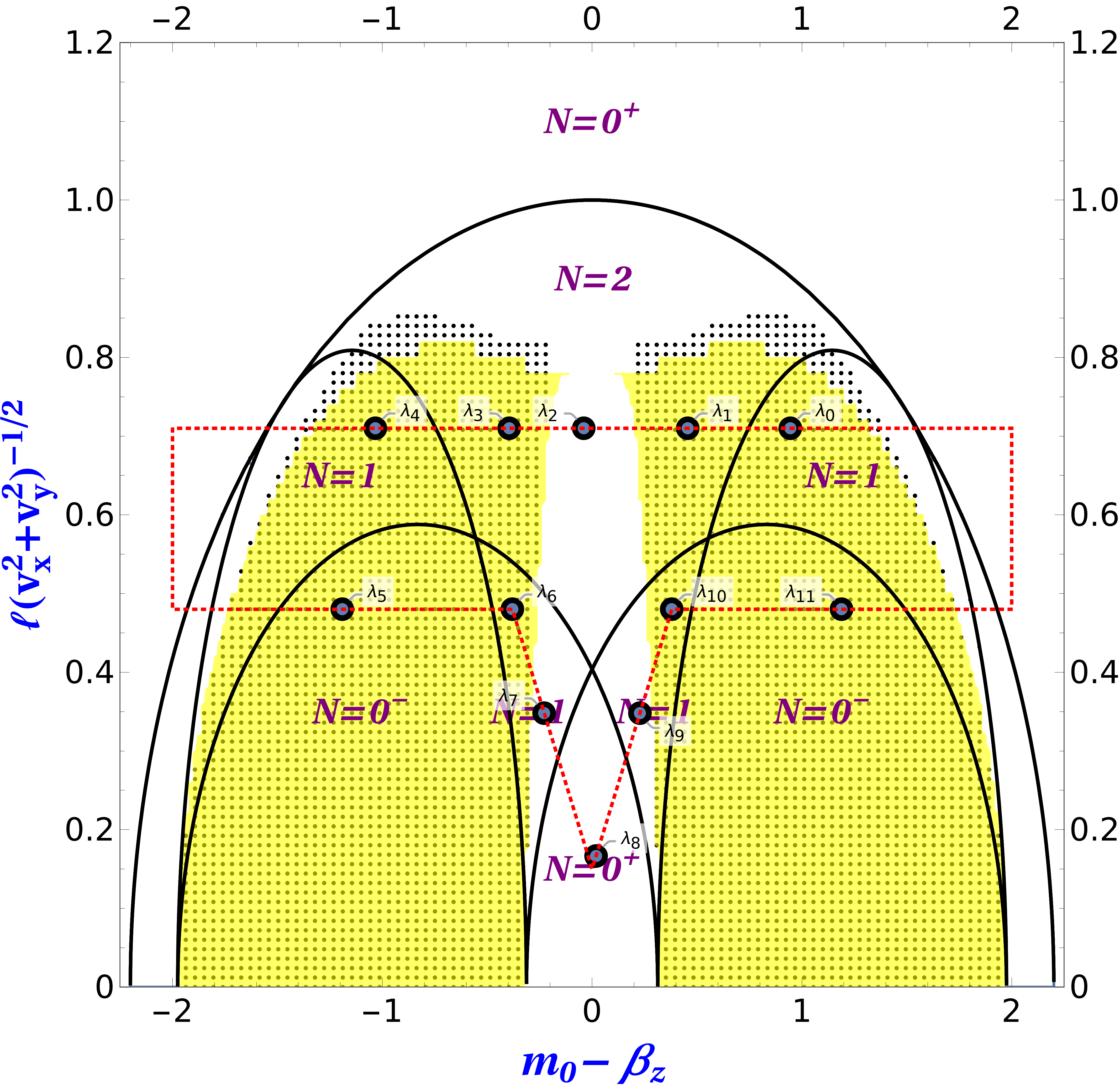}

}\\

\subfloat[]{\includegraphics[clip,width=0.9\linewidth]{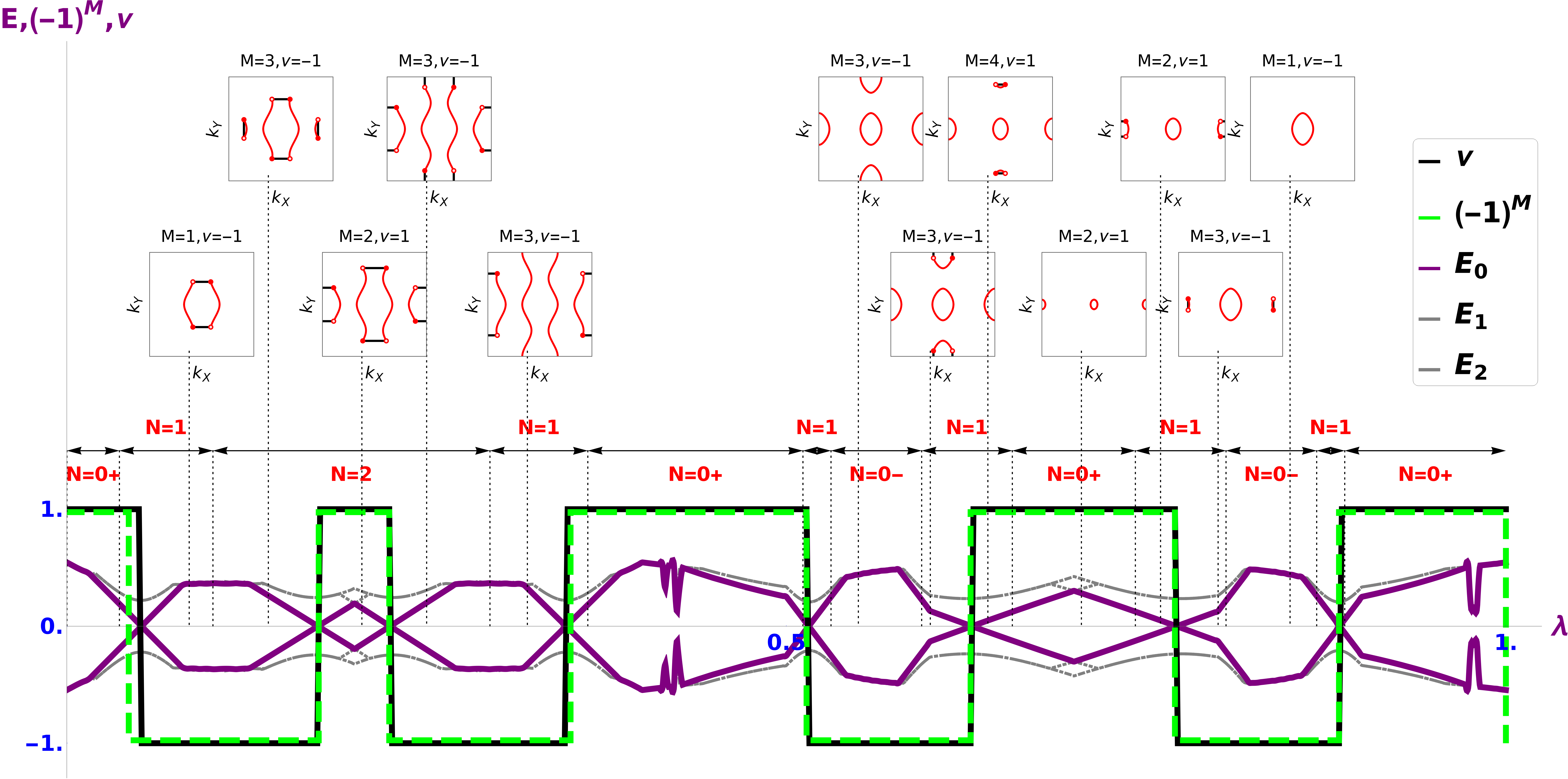}

}\caption{\jcaption{(a) Representative points and the path for which results
are presented in (b). (b) Lowest few energies, predicted topological
invariant $(-1)^{M}$ and the computed invariant $\nu$ along the
path denoted in (a) parameterized by $\lambda$. The insets show the
FGSs at each representative point. The predicted and computed results
show excellent agreement and each phase transition is accompanied
by a level crossing at zero energy.}\label{fig:Level-crossing-1}}
\end{figure}

In this section, we select representative points on the phase diagram
in Fig. 2 of the main paper, explicitly compare the prediction (Equation
1 of the main paper) and the result of calculating the Pfaffian-based
invariant \cite{Kitaev2000}, and show that topological phase transitions
are accompanied by level crossings in the vortex as expected. The
results are shown in Fig. \ref{fig:Level-crossing-1}. All the data
points in Fig. 2 of the main paper were obtained using the same Pfaffian-based
invariant \cite{Kitaev2000}.

%
%
%
%
%

\end{document}